\documentclass[twocolumn,showpacs,superscriptaddress,amsmath,amssymb,prl]{revtex4-1}
\usepackage[colorlinks,linkcolor=blue,anchorcolor=blue,citecolor=blue,urlcolor=blue]{hyperref}
\usepackage{graphicx}
\usepackage{epsfig}
\usepackage{dcolumn}
\usepackage{bm}
\usepackage{overpic}
\usepackage{color}
\usepackage{multirow}
\usepackage{subfigure}

\usepackage{verbatim}
\usepackage{cleveref}

\newcommand\EPP{e^+e^-\to\eta\pi^{+}\pi^-}
\newcommand\RHOETA{e^+e^-\to\rho\eta}
\newcommand\API{e^+e^-\to a_2(1320)\pi}

\begin{document}
\newcommand{\ks}{K_{S}^{0}}
\newcommand{\EP}{e^{+}}
\newcommand{\EM}{e^{-}}
\newcommand{\epm}{e^{\pm}}
\newcommand{\vpho}{\gamma^{\ast}}
\newcommand{\qqbar}{q\bar{q}}

\newcommand{\ee}{e^{+}e^{-}}
\newcommand{\mm}{\mu^{+}\mu^{-}}
\newcommand{\alfs}{\alpha_{s}}
\newcommand{\alfmz}{\alpha(M_{Z}^{2})}
\newcommand{\amu}{a_{\mu}}
\newcommand{\Lam}{\Lambda_{c}}
\newcommand{\lam}{\Lambda_{c}^{+}}
\newcommand{\lambar}{\bar{\Lambda}_{c}^{-}}
\newcommand{\Lambdac}{\Lambda_{c}}
\newcommand{\mbc}{M_{BC}}
\newcommand{\dele}{\Delta E}
\newcommand{\ebm}{E_{\textmd{beam}}}
\newcommand{\ecm}{E_{\textmd{c.m.}}}
\newcommand{\pbm}{p_{\textmd{beam}}}
\newcommand{\MuMu}{\mu\mu}
\newcommand{\mumu}{\mu\mu}
\newcommand{\tata}{\tau^{+}\tau^{-}}
\newcommand{\pipi}{\pi^{+}\pi^{-}}
\newcommand{\gaga}{\gamma\gamma}
\newcommand{\twopho}{\ee+X}
\newcommand{\sqs}{\sqrt{s}}
\newcommand{\sqsp}{\sqrt{s^{\prime}}}
\newcommand{\da}{\Delta\alpha}
\newcommand{\das}{\Delta\alpha(s)}
\newcommand{\dimu}{\ee \ra \mumu}
\newcommand{\dedx}{\textmd{d}E/\textmd{d}x}
\newcommand{\chip}{\chi_{\textmd{Prob}}}
\newcommand{\chiP}{\chi_{p}}
\newcommand{\evz}{V_{z}^{\textmd{evt}}}
\newcommand{\evzloose}{V_{z,\textmd{loose}}^{\textmd{evt}}}
\newcommand{\avz}{V_{z}^{\textmd{ave}}}
\newcommand{\Ngd}{N_{\textmd{good}}}
\newcommand{\Ncru}{N_{\textmd{crude}}}
\newcommand{\pio}{\pi^{0}}
\newcommand{\rpid}{r_{\textmd{PID}}}

\newcommand{\Nhxobs}{N_{h+X}^{\textmd{obs}}}
\newcommand{\Nhobs}{N_{h}^{\textmd{obs}}}
\newcommand{\Npioxobs}{N_{\pi^{0}+X}^{\textmd{obs}}}
\newcommand{\Nksxobs}{N_{\ks+X}^{\textmd{obs}}}
\newcommand{\Npioobs}{N_{\pi^{0}}^{\textmd{obs}}}
\newcommand{\Nksobs}{N_{\ks}^{\textmd{obs}}}
\newcommand{\Nhxtru}{N_{h+X}^{\textmd{tru}}}
\newcommand{\Nhtru}{N_{h}^{\textmd{tru}}}
\newcommand{\Npiotru}{N_{\pi^{0}}^{\textmd{tru}}}
\newcommand{\Nkstru}{N_{\ks}^{\textmd{tru}}}
\newcommand{\Nbarhxobs}{\bar{N}_{h+X}^{\textmd{obs}}}
\newcommand{\Nbarhobs}{\bar{N}_{h}^{\textmd{obs}}}
\newcommand{\Nbarpioobs}{\bar{N}_{\pi^{0}}^{\textmd{obs}}}
\newcommand{\Nbarksobs}{\bar{N}_{\ks}^{\textmd{obs}}}
\newcommand{\Nbarhxtru}{\bar{N}_{h+X}^{\textmd{tru}}}
\newcommand{\Nbarhtru}{\bar{N}_{h}^{\textmd{tru}}}
\newcommand{\Nbarpiotru}{\bar{N}_{\pi^{0}}^{\textmd{tru}}}
\newcommand{\Nbarkstru}{\bar{N}_{\ks}^{\textmd{tru}}}

\newcommand{\Nhadtot}{N_{\textmd{had}}^{\textmd{tot}}}
\newcommand{\Nhadobs}{N_{\textmd{had}}^{\textmd{obs}}}
\newcommand{\Nbarhadobs}{\bar{N}_{\textmd{had}}^{\textmd{obs}}}
\newcommand{\Nhadtru}{N_{\textmd{had}}^{\textmd{tru}}}
\newcommand{\Nbarhadtru}{\bar{N}_{\textmd{had}}^{\textmd{tru}}}
\newcommand{\Nhadphy}{N_{\textmd{had}}}

\newcommand{\cshadobs}{\sigma_{\textmd{had}}^{\textmd{obs}}}
\newcommand{\effhad}{\vap_{\textmd{had}}}
\newcommand{\efftrg}{\vap_{\textmd{trig}}}
\newcommand{\lint}{\mathcal{L}_{\textmd{int}}}
\newcommand{\Nbkg}{N_{\textmd{bkg}}}
\newcommand{\NbkgTot}{N_{\textrm{bkg}}^{\textrm{Tot}}}
\newcommand{\csbkg}{\sigma_{\textmd{bkg}}}
\newcommand{\Nmcsur}{N_{\textmd{MC}}^{\textmd{sur}}}
\newcommand{\Nmcsurori}{N_{\textmd{MC}}^{\textmd{sur,nom.}}}
\newcommand{\Nmcsurwtd}{N_{\textmd{MC}}^{\textmd{sur,wtd.}}}
\newcommand{\Nmcgen}{N_{\textmd{MC}}^{\textmd{gen}}}
\newcommand{\vap}{\varepsilon}
\newcommand{\chisq}{\chi^{2}}
\newcommand{\cshadphy}{\sigma_{\textmd{had}}^{\textmd{phy}}}
\newcommand{\cshadtot}{\sigma_{\textmd{had}}^{\textmd{tot}}}
\newcommand{\cshadborn}{\sigma_{\textmd{had}}^{0}}
\newcommand{\cshadborncon}{\sigma_{\textmd{con}}^{0}}
\newcommand{\cshadbornres}{\sigma_{\textmd{res}}^{0}}
\newcommand{\csdimuborn}{\sigma_{\mu\mu}^{0}}
\newcommand{\rpqcd}{R_{\textmd{pQCD}}}
\newcommand{\Nprod}{N_{\textmd{prod}}}
\newcommand{\Nhadnet}{N_{\textrm{had}}^{\textrm{net}}}
\newcommand{\Delrel}{\Delta_{\textrm{rel}}}

\newcommand{\fourpionchg}{\pipi\pipi}
\newcommand{\fourpionneu}{\pipi\pi^{0}\pi^{0}}
\newcommand{\sixpionchg}{3(\pipi)}
\newcommand{\thrpionneu}{\pipi\pi^{0}}
\newcommand{\twopionchg}{\pipi}

\newcommand{\Nsurnpion}{N_{\textmd{sur}}^{n\pi}}
\newcommand{\Ngennpion}{N_{\textmd{gen}}^{n\pi}}
\newcommand{\Ngentot}{N_{\textmd{gen}}^{\textmd{tot}}}
\newcommand{\effincnpion}{\vap_{n\pi}^{\textmd{inc}}}
\newcommand{\effincnpionp}{\vap_{n\pi}^{\textmd{inc},\prime}}
\newcommand{\effincnonnpion}{\vap_{\textmd{non}-n\pi}^{\textmd{inc}}}
\newcommand{\effexcnpion}{\vap_{n\pi}^{\textmd{exc}}}
\newcommand{\fracnpion}{f_{n\pi}}
\newcommand{\fracnpionp}{f_{n\pi}^{\prime}}
\newcommand{\fracnonnpion}{f_{\textmd{non}-n\pi}}

\newcommand{\Nsurtwopion}{N_{\textmd{sur}}^{2\pi}}
\newcommand{\Ngentwopion}{N_{\textmd{gen}}^{2\pi}}
\newcommand{\effinctwopion}{\vap_{2\pi}^{\textmd{inc}}}
\newcommand{\effinctwopionp}{\vap_{2\pi}^{\textmd{inc},\prime}}
\newcommand{\effincnontwopion}{\vap_{\textmd{non}-2\pi}^{\textmd{inc}}}
\newcommand{\effexctwopion}{\vap_{2\pi}^{\textmd{exc}}}
\newcommand{\fractwopion}{f_{2\pi}}
\newcommand{\fractwopionp}{f_{2\pi}^{\prime}}
\newcommand{\fracnontwopion}{f_{\textmd{non}-2\pi}}

\newcommand{\Nsurthrpion}{N_{\textmd{sur}}^{3\pi}}
\newcommand{\Ngenthrpion}{N_{\textmd{gen}}^{3\pi}}
\newcommand{\effincthrpion}{\vap_{3\pi}^{\textmd{inc}}}
\newcommand{\effincthrpionp}{\vap_{3\pi}^{\textmd{inc},\prime}}
\newcommand{\effincnonthrpion}{\vap_{\textmd{non}-3\pi}^{\textmd{inc}}}
\newcommand{\effexcthrpion}{\vap_{3\pi}^{\textmd{exc}}}
\newcommand{\fracthrpion}{f_{3\pi}}
\newcommand{\fracthrpionp}{f_{3\pi}^{\prime}}
\newcommand{\fracnonthrpion}{f_{\textmd{non}-3\pi}}

\newcommand{\Nsurfourpion}{N_{\textmd{sur}}^{4\pi}}
\newcommand{\Ngenfourpion}{N_{\textmd{gen}}^{4\pi}}
\newcommand{\effincfourpion}{\vap_{4\pi}^{\textmd{inc}}}
\newcommand{\effincfourpionp}{\vap_{4\pi}^{\textmd{inc},\prime}}
\newcommand{\effincnonfourpion}{\vap_{\textmd{non}-4\pi}^{\textmd{inc}}}
\newcommand{\effexcfourpion}{\vap_{4\pi}^{\textmd{exc}}}
\newcommand{\fracfourpion}{f_{4\pi}}
\newcommand{\fracfourpionp}{f_{4\pi}^{\prime}}
\newcommand{\fracnonfourpion}{f_{\textmd{non}-4\pi}}

\newcommand{\Npionprod}{N_{\textmd{prod}}^{4\pi}}
\newcommand{\Ndatasur}{N_{\textmd{data}}^{\textmd{sur}}}
\newcommand{\Nobspion}{N_{\textmd{obs}}^{4\pi}}
\newcommand{\Nhadprod}{N_{\textmd{prod}}^{\textmd{had}}}
\newcommand{\sigmaobs}{\sigma_{\textmd{obs}}}
\newcommand{\effhadp}{\vap_{\textmd{had}}^{\prime}}

\newcommand{\effpion}{\vap_{4\pi}}
\newcommand{\effexcpion}{\vap_{4\pi}^{\textmd{exc}}}
\newcommand{\effincpion}{\vap_{4\pi}^{\textmd{inc}}}
\newcommand{\effincpionI}{\vap_{4\pi}^{\textmd{inc},1}}
\newcommand{\effincpionII}{\vap_{4\pi}^{\textmd{inc},2}}
\newcommand{\effincpionp}{\vap_{4\pi}^{\textmd{inc},\prime}}
\newcommand{\effincremain}{\vap_{\textmd{non}-n\pi}^{\textmd{inc}}}

\newcommand{\fracpion}{f_{4\pi}}
\newcommand{\fracnonpion}{f_{\textmd{non}-4\pi}}
\newcommand{\fracnonpionp}{f_{\textmd{non}-4\pi}^{\prime}}
\newcommand{\fracpionII}{f_{4\pi}^{2}}
\newcommand{\fracpionp}{f_{4\pi}^{\prime}}
\newcommand{\reladiff}{\Delta_{\textmd{rel}}}

\newcommand{\Nsursixpion}{N_{\textmd{sur}}^{6\pi}}
\newcommand{\Ngensixpion}{N_{\textmd{gen}}^{6\pi}}
\newcommand{\effincsixpion}{\vap_{6\pi}^{\textmd{inc}}}
\newcommand{\fracsixpion}{f_{6\pi}}

\newcommand{\etot}{E_{\textmd{tot}}}
\newcommand{\ptot}{p_{\textmd{tot}}}
\newcommand{\plab}{p_{\textmd{Lab}}}
\newcommand{\mpiOI}{M(\pi^{0}_{1})}
\newcommand{\mpiOII}{M(\pi^{0}_{2})}

\newcommand{\widtheeoi}{\varGamma^{\textmd{ee}}_{0,i}}
\newcommand{\widtheeoj}{\varGamma^{\textmd{ee}}_{0,j}}
\newcommand{\widtheeo}{\varGamma^{\textmd{ee}}_{0}}
\newcommand{\widthee}{\varGamma^{\textmd{ee}}}
\newcommand{\widtheeexpi}{\varGamma^{\textmd{ee}}_{\textmd{exp},i}}
\newcommand{\widtheeexp}{\varGamma^{\textmd{ee}}_{\textmd{exp}}}
\newcommand{\widthtoti}{\varGamma^{\textmd{tot}}_{i}}
\newcommand{\widthtot}{\varGamma^{\textmd{tot}}}

\newcommand{\vpqed}{\Pi_{\textmd{QED}}}
\newcommand{\vpqcd}{\Pi_{\textmd{QCD}}}
\newcommand{\vpcon}{\Pi_{\textmd{con}}}
\newcommand{\vpres}{\Pi_{\textmd{res}}}
\newcommand{\vpo}{\Pi_{0}}
\newcommand{\rcon}{R_{\textmd{con}}}
\newcommand{\rres}{R_{\textmd{res}}}
\newcommand{\rexp}{R_{\textmd{exp}}}

\newcommand{\delvert}{\delta_{\textmd{vert}}}
\newcommand{\delvp}{\delta_{\textmd{vac}}}
\newcommand{\delbrem}{\delta_{\gamma}}
\newcommand{\delobs}{\delta_{\textmd{obs}}}
\newcommand{\radiatorsf}{F_{\textmd{SF}}}
\newcommand{\radiatorfd}{F_{\textmd{FD}}}
\newcommand{\DelFD}{\Delta_{\textmd{FD}}}
\newcommand{\DelFDCal}{\Delta_{\textmd{cal}}}
\newcommand{\DelFDcs}{\Delta_{\sigma}}
\newcommand{\DelFDvp}{\Delta_{\textmd{vp}}}

\newcommand{\costh}{\cos\theta}
\newcommand{\costhIprg}{\cos\theta_{\textmd{1prg}}}
\newcommand{\costhIIprg}{\cos\theta_{\textmd{2prg}}}
\newcommand{\costhIIIprg}{\cos\theta_{\textmd{3prg}}}
\newcommand{\costhIVprg}{\cos\theta_{\textmd{4prg}}}
\newcommand{\costhrestprg}{\cos\theta_{\textmd{restprg}}}
\newcommand{\emce}{E^{\textmd{ctrk.}}_{\textmd{emc}}}
\newcommand{\emceIprg}{E^{\textmd{ctrk.}}_{\textmd{emc,1prg}}}
\newcommand{\emceIIprg}{E^{\textmd{ctrk.}}_{\textmd{emc,2prg}}}
\newcommand{\emceIIIprg}{E^{\textmd{ctrk.}}_{\textmd{emc,3prg}}}
\newcommand{\emceIVprg}{E^{\textmd{ctrk.}}_{\textmd{emc,4prg}}}
\newcommand{\emcerestprg}{E^{\textmd{ctrk.}}_{\textmd{emc,restprg}}}
\newcommand{\isocosth}{\cos\theta_{\textmd{iso}}}
\newcommand{\isocosthIprg}{\cos\theta_{\textmd{iso,1prg}}}
\newcommand{\isocosthIIprg}{\cos\theta_{\textmd{iso,2prg}}}
\newcommand{\isocosthIIIprg}{\cos\theta_{\textmd{iso,3prg}}}
\newcommand{\isocosthIVprg}{\cos\theta_{\textmd{iso,4prg}}}
\newcommand{\isocosthrestprg}{\cos\theta_{\textmd{iso,restprg}}}
\newcommand{\eop}{E/P}
\newcommand{\eopIprg}{E/P_{\textmd{1prg}}}
\newcommand{\eopIIprg}{E/P_{\textmd{2prg}}}
\newcommand{\eopIIIprg}{E/P_{\textmd{3prg}}}
\newcommand{\eopIVprg}{E/P_{\textmd{4prg}}}
\newcommand{\eoprestprg}{E/P_{\textmd{restprg}}}
\newcommand{\nisogam}{N_{\textmd{isogam}}}
\newcommand{\nisogamIprg}{N_{\textmd{isogam,1prg}}}
\newcommand{\nisogamIIprg}{N_{\textmd{isogam,2prg}}}
\newcommand{\nisogamIIIprg}{N_{\textmd{isogam,3prg}}}
\newcommand{\nisogamIVprg}{N_{\textmd{isogam,4prg}}}
\newcommand{\nisogamrestprg}{N_{\textmd{isogam,restprg}}}
\newcommand{\ptrk}{p_{\textmd{ctrk}}}
\newcommand{\pIprg}{p^{\textmd{ctrk}}_{\textmd{1prg}}}
\newcommand{\pIIprg}{p^{\textmd{ctrk}}_{\textmd{2prg}}}
\newcommand{\pIIIprg}{p^{\textmd{ctrk}}_{\textmd{3prg}}}
\newcommand{\pIVprg}{p^{\textmd{ctrk}}_{\textmd{4prg}}}
\newcommand{\prestprg}{p^{\textmd{ctrk}}_{\textmd{restprg}}}
\newcommand{\tote}{E_{\textmd{vis.}}}
\newcommand{\totevte}{E_{\textmd{tot.}}}
\newcommand{\totevteIprg}{E_{\textmd{tot.}}^{\textmd{1prg}}}
\newcommand{\balanceIprg}{\textmd{Balance}}
\newcommand{\ngamma}{N_{\gamma}}
\newcommand{\ngammaIprg}{N_{\gamma,\textmd{1prg}}}
\newcommand{\ngammaIIprg}{N_{\gamma,\textmd{2prg}}}
\newcommand{\ngammaIIIprg}{N_{\gamma,\textmd{3prg}}}
\newcommand{\ngammaIVprg}{N_{\gamma,\textmd{4prg}}}
\newcommand{\ngammarestprg}{N_{\gamma,\textmd{restprg}}}
\newcommand{\ngoodwt}{N_{\textmd{good}}^{\textmd{Wt}}}
\newcommand{\ngood}{N_{\textmd{good}}}
\newcommand{\npiO}{N_{\pi^{0}}}
\newcommand{\npiOIprg}{N_{\pi^{0}}^{\textmd{1prg}}}
\newcommand{\npiOIIprg}{N_{\pi^{0}}^{\textmd{2prg}}}
\newcommand{\npp}{N_{p}}
\newcommand{\npm}{N_{\bar{p}}}
\newcommand{\nkp}{N_{K^{+}}}
\newcommand{\nkm}{N_{K^{-}}}
\newcommand{\npip}{N_{\pi^{+}}}
\newcommand{\npim}{N_{\pi^{-}}}
\newcommand{\ppp}{P(p^{+})}
\newcommand{\ppm}{P(\bar{p}^{-})}
\newcommand{\pkp}{p(K^{+})}
\newcommand{\pkm}{p(K^{-})}
\newcommand{\ppip}{P(\pi^{+})}
\newcommand{\ppim}{P(\pi^{-})}
\newcommand{\ppiO}{P(\pi^{0})}
\newcommand{\mpiO}{M(\pi^{0})}
\newcommand{\mks}{M(K^{0}_{s})}
\newcommand{\pks}{p_{K^{0}_{s}}}
\newcommand{\mphi}{M(\phi)}
\newcommand{\pphi}{p_{\phi}}
\newcommand{\mIIgam}{M(\gamma\gamma)}
\newcommand{\mIIgamIprg}{M(\gamma\gamma)^{\textmd{1prg}}}
\newcommand{\pIIgam}{p_{\gamma\gamma}}
\newcommand{\mlambda}{M(\Lambda)}
\newcommand{\plambda}{p_{\Lambda}}
\newcommand{\mdO}{M(D^{0})}
\newcommand{\pdO}{p_{D^{0}}}
\newcommand{\mdstarO}{M(D^{\ast 0})}
\newcommand{\pdstarO}{p_{D^{\ast 0}}}
\newcommand{\mdp}{M(D^{\pm})}
\newcommand{\pdp}{p_{D^{\pm}}}
\newcommand{\mdstarp}{M(D^{\ast\pm})}
\newcommand{\pdstarp}{p_{D^{\ast\pm}}}
\newcommand{\mds}{M(D_{s}^{\pm})}
\newcommand{\pds}{p_{D_{s}^{\pm}}}
\newcommand{\mdstars}{M(D_{s}^{\ast\pm})}
\newcommand{\pdstars}{p_{D_{s}^{\ast\pm}}}
\newcommand{\Vr}{V_{r}}
\newcommand{\Vz}{V_{z}}

\newcommand{\gev}{\mathrm{GeV}}
\newcommand{\mev}{\mathrm{MeV}}
\newcommand{\mevcc}{\mathrm{MeV}/c^{2}}
\newcommand{\gevc}{\mathrm{GeV}/c}
\newcommand{\gevcc}{\mathrm{GeV}/c^2}

\newcommand{\nchg}{N_{\textmd{chg}}}
\newcommand{\eff}{\vap}

\newcommand{\critecm}{1.780}

\newcommand{\ENERGYAT}{4575.5}
\newcommand{\ENERGYBT}{4575.5}
\newcommand{\ENERGYCT}{4575.5}
\newcommand{\ENERGYDT}{4575.5}
\newcommand{\ksdecay}{\ks\ra\pi^{+}\pi^{-}}
\newcommand{\phidecay}{\phi\ra K^{+}K^{-}}
\newcommand{\piOdecay}{\pi^{0}\ra\gamma\gamma}
\newcommand{\Lambdadecay}{\Lambda\ra p\pi^{-}}
\newcommand{\DOdecay}{D^{0}\ra K^{-}\pi^{+}}
\newcommand{\DStarOdecay}{D^{\ast0}\ra D^{0}\pi^{0}}
\newcommand{\Dpdecay}{D^{+}\ra K^{+}\pi^{+}\pi^{-}}
\newcommand{\DStarpdecay}{D^{\ast+}\ra D^{0}\pi^{+}}
\newcommand{\Dsdecay}{D^{+}_{s}\ra K^{+}K^{-}\pi^{+}}
\newcommand{\DStarsdecay}{D^{\ast+}_{s}\ra D^{+}_{s}\gamma}


\title{\boldmath \textbf{Measurement of the cross sections for $e^+e^-\to\eta\pi^+\pi^-$ at center-of-mass energies between 2.00 and 3.08~GeV} }

\author{
M.~Ablikim$^{1}$, M.~N.~Achasov$^{4,b}$, P.~Adlarson$^{75}$, X.~C.~Ai$^{80}$, R.~Aliberti$^{35}$, A.~Amoroso$^{74A,74C}$, M.~R.~An$^{39}$, Q.~An$^{71,58}$, Y.~Bai$^{57}$, O.~Bakina$^{36}$
% ...
\\
\vspace{0.2cm}
(BESIII Collaboration)\\
\vspace{0.2cm} {\it
$^{1}$ Institute of High Energy Physics, Beijing 100049, People's Republic of China\\
% ...
$^{4}$ Budker Institute of Nuclear Physics SB RAS (BINP), Novosibirsk 630090, Russia\\
% ...
$^{35}$ Johannes Gutenberg University of Mainz, Johann-Joachim-Becher-Weg 45, D-55099 Mainz, Germany\\
$^{36}$ Joint Institute for Nuclear Research, 141980 Dubna, Moscow region, Russia\\
% ...
$^{39}$ Liaoning Normal University, Dalian 116029, People's Republic of China\\
% ...
$^{57}$ Southeast University, Nanjing 211100, People's Republic of China\\
$^{58}$ State Key Laboratory of Particle Detection and Electronics, Beijing 100049, Hefei 230026, People's Republic of China\\
% ...
$^{71}$ University of Science and Technology of China, Hefei 230026, People's Republic of China\\
% ...
$^{74}$ University of Turin and INFN, (A)University of Turin, I-10125, Turin, Italy; (B)University of Eastern Piedmont, I-15121, Alessandria, Italy; (C)INFN, I-10125, Turin, Italy\\
$^{75}$ Uppsala University, Box 516, SE-75120 Uppsala, Sweden\\
% ...
$^{80}$ Zhengzhou University, Zhengzhou 450001, People's Republic of China\\
\vspace{0.2cm}
% ...
$^{b}$ Also at the Novosibirsk State University, Novosibirsk, 630090, Russia\\
% ...
}
}

\date{\today}

\begin{abstract}
Using data samples collected at center-of-mass energies between 2.000 and 3.080~GeV
with the BESIII detector operating at the BEPCII collider,
a partial-wave analysis is performed on the process $\EPP$.
In addition to the dominant $\RHOETA$ component, the $\API$ process
is also sizeable, contributing up to 24\% of the total reaction.
The measured cross sections of the process $\EPP$ are systematically higher than those of BaBar
by more than $3\sigma$ at center-of-mass energies between 2.000 and 2.300~GeV.
In the cross section lineshape for $\API$, a resonant structure is observed with a significance of $5.5\sigma$, with $M=(2044\pm31\pm4)$~MeV/$c^2$, $\Gamma=(163\pm69\pm24)$~MeV
and $\mathcal{B_{R}}\cdot\Gamma_{e^+e^-}^{R}=(34.6\pm17.1\pm6.0)$~eV or
$(137.1\pm73.3\pm2.1)$~eV.
In the cross section lineshape for $\RHOETA$, an evidence of a dip structure around 2180~MeV/$c^2$ is observed with statistical significance of $3.0\sigma$.

\end{abstract}

\maketitle

Determining the hadronic contribution to the muon anomalous magnetic moment ($g_\mu-2$) is
currently a high priority in hadronic physics.
The latest measurement of $(g_\mu-2)$ from Fermilab increased the tension between experiments and theories to
a $5.0\sigma$ discrepancy~\cite{Muong-2:2023cdq,Aoyama:2020ynm},
while recent lattice gauge theory predictions reduced this discrepancy~\cite{Borsanyi:2020mff}.
The Standard Model (SM) calculation of ($g_\mu-2$) requires inputs from experimental $e^+e^-$ hadronic cross section
data to account for the hadronic vacuum polarization (HVP) term.
An updated SM calculation that considers all available hadronic data
will likely yield a less significant discrepancy with experiment~\cite{Muong-2:2023cdq}.
In the calculation, a sum of exclusive states must be used, in which the $\EPP$ process has a sizeable contribution~\cite{Davier:2017zfy,Jegerlehner:2018zrj,Keshavarzi:2018mgv}.
Improved measurements of this process are therefore important to improve the reliability of the ($g_\mu-2$) calculation.

The process $\EPP$ has previously been studied at energies from threshold to 3.5~GeV by several experiments: DM1~\cite{CORDIER198013}, ND~\cite{DRUZHININ1986115}, DM2~\cite{DM2:1988uai}, CMD-2~\cite{CMD-2:2000mlo}, SND~\cite{Achasov:2010zzd,SND:2014rfi,Achasov:2017kqm}, BaBar~\cite{BaBar:2007qju,BaBar:2018erh}, and CMD-3~\cite{Gribanov:2019qgw}.
In previous studies, the $\EPP$ final state was simulated using only the $\RHOETA$ hadronic intermediate state,
while other intermediate processes
such as $\API$
were not considered in the determination of the efficiency.

The two-body processes $\RHOETA$ and $\API$ are important for the spectroscopy of the excited $\rho-$like states, including the $\rho(2000)$, $\rho(2150)$, and $\rho(2270)$.
The $\rho(2000)$ was found in $p\bar{p}$ collisions~\cite{Hasan:1994he,Anisovich:2002su}, and has been explained as a radial excitation of the $\rho(1700)$~\cite{Anisovich:2002su} and as a mixed state with a $^3D_1$ component~\cite{Bugg:2004xu,Li:2021qgz}.
Ref.~\cite{Li:2021qgz} indicates that $\API$ is a good channel for investigating $\rho(2000)$.
The $\rho(2150)$ was initially regarded as a $2^3D_1$ state~\cite{Godfrey:1985xj} but was later considered to be a $4^3S_1$ state~\cite{Bugg:2012yt,Masjuan:2013xta,He:2013ttg,Li:2021qgz}.
The $\rho(2150)$ has been widely studied in $e^+e^-$, $p\bar{p}$, $s-$channel $N\bar{N}$, and $\pi p$ experiments~\cite{ParticleDataGroup:2022pth},
but inconsistencies in the measured masses and widths make the $\rho(2150)$ more controversial.
The $\rho(2270)$ was first observed in photoproduction~\cite{OmegaPhoton:1985vyt} and categorized as a $3^3D_1$ state~\cite{Bugg:2012yt,He:2013ttg}.
Up to now there are no published results on the production of the $\rho(2270)$ in $e^+e^-$ collision experiments.

In this Letter, we present a partial-wave analysis (PWA) of the process $\EPP$ using data collected with the BESIII detector.
There are 19 datasets used in this analysis, with center-of-mass (c.m.) energies from
2.00 to 3.08~GeV and a total integrated luminosity of 648~$\rm pb^{-1}$.
The charge-conjugated processes are included by default in the following discussions.

The BESIII detector has a geometrical acceptance of 93\% of the full solid angle and consists of four main components:
(i) a small cell, helium-based multi-layer drift chamber (MDC),
(ii) a time-of-flight system made from two layers of plastic scintillator,
(iii) an electromagnetic calorimeter (EMC) made of CsI(Tl) crystals,
and (iv) a resistive plate chamber–based muon chamber.
More details of the design and performance of the BESIII detector can be found in Ref.~\cite{BESIII:2009fln}.
The {\sc geant4}~\cite{GEANT4:2002zbu} based Monte Carlo (MC) simulation of the full detector
is performed to optimize the event selection criteria,
to understand potential backgrounds, and to determine the detection efficiency.
The signal MC samples are generated with {\sc conexc}~\cite{Ping:2016pms}, which incorporates a higher-order initial state radiation (ISR) correction. The subsequent decay $\eta\to\gamma\gamma$ is simulated by {\sc besevtgen}~\cite{Lange:2001uf,Ping:2008zz}.
Inclusive MC samples equivalent to twice the data sets at $\sqrt{s}$= 2.1250, 2.396 and 2.900 GeV are generated, respectively, to study the backgrounds.
The $e^+e^-\to(\gamma) e^+e^-, (\gamma)\mu^+\mu^-$ and $\gamma\gamma$ events are simulated with the {\sc Babayaga} generator~\cite{Balossini:2006wc},
while $e^+e^-\to$~hadrons events are generated with a hybrid generator~\cite{Ping:2016pms}.

To identify $\EPP$ candidates, final states with only one oppositely charged pion pair
and at least two photons are selected.
The selection criteria for charged tracks, particle identification (PID), and photons are the same as described in Ref.~\cite{BESIII:2020kpr}.
A vertex fit is imposed on the selected charged tracks to ensure that they originate from the same interaction point.
The $\eta$ meson is reconstructed with $\eta\to\gamma\gamma$.
To suppress background due to the mis-combination of photons, $\cos\theta_\gamma$ is required to be less than 0.95, where $\theta_{\gamma}$ is the polar angle of one photon in the helicity frame of the $\eta$ meson.
To further suppress the background,
a four-constraint (4C) kinematic fit imposing energy-momentum conservation is employed,
and $\chi^2_{4{\rm C}}<100$ is required.
The Bhabha events are removed with the requirement of $E/p<0.8$, where $E$ is the deposited energy in the EMC and $p$ is the momentum measured by the MDC for the charged pion.
Based on an analysis of the inclusive MC events, the processes $e^+e^-\to\pi^+\pi^-\pi^0\pi^0$ and $e^+e^-\to\pi^+\pi^-\pi^0$ are found to be the dominant backgrounds, and no peaking background is observed.
Signal candidates are required to be within the $\eta$ mass signal region, which is defined as [0.523, 0.573] $\gevcc$ on the photon pair invariant mass.
The $M(\gamma\gamma)$ distribution at $\sqrt{s}=2.125$ is shown in Fig.~\ref{fit_eta2125},
and Fig.1 of the Supplemental Material~\cite{suppstuff} shows the $M(\gamma\gamma)$ at $\sqrt{s}=2.396$ and $\sqrt{s}=2.900$~GeV.
The events in the $\eta$ mass sideband regions, which are defined as [0.478, 0.503] and [0.593, 0.618] $\gevcc$, are used to estimate the background.
The signal purities after sideband subtraction for different c.m.\ energies are listed in Table~I of the Supplemental Material~\cite{suppstuff}.

\begin{figure}[h]
	\centering
	\includegraphics[width=8.5cm]{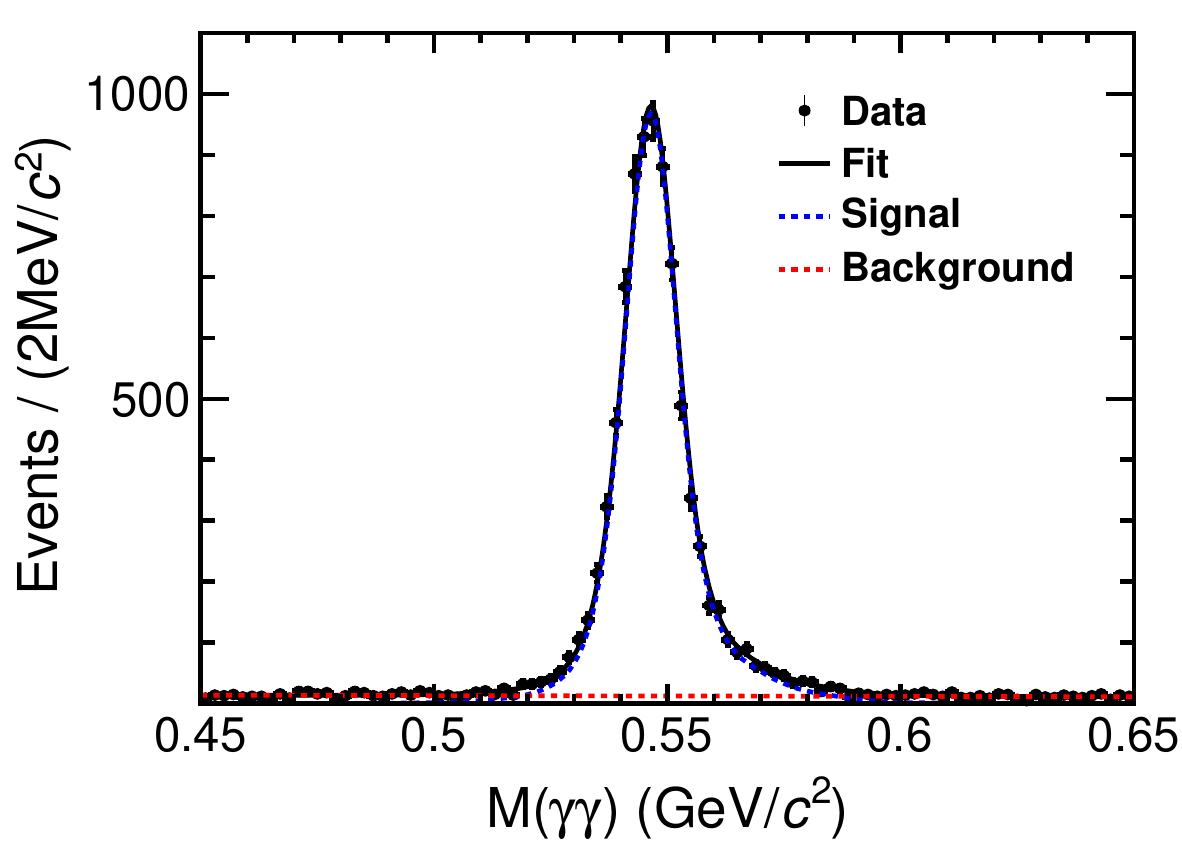}
    \vspace*{-0.3cm}
	\caption{ \small Fit to the $M(\gamma\gamma)$ distribution at $\sqrt{s}=2.125$~GeV. Dots with error bars are data, blue dashed line is the signal shape, red dashed line is the continuum background, the black line gives the total fit result. }
	\label{fit_eta2125}
\end{figure}

To obtain a reliable efficiency for the process $\EPP$ and to extract the contributions from the intermediate processes,
a PWA based on the GPUPWA framework~\cite{Berger:2010zza}
is performed on the surviving candidate events.
The quasi two-body decay amplitudes in the sequential decay
processes $e^+e^-\to\eta X$, $X\to\pi^+\pi^-$ and $e^+e^-\to\pi^+(\pi^-)X$, $X\to\pi^-(\pi^+)\eta$
are constructed using covariant tensor amplitudes~\cite{Zou:2002ar}.
The intermediate state $X$ is parameterized by relativistic Breit-Wigner (BW) functions with constant widths,
except for the wide $\rho(770)$,
which is described by the Gounaris-Sakurai model~\cite{Gounaris:1968mw}.
The complex coefficients of the amplitudes (relative magnitudes and phases) of the individual intermediate processes
are determined by performing an unbinned maximum likelihood fit using MINUIT~\cite{James:1975dr}.
The probability for the observed events is characterized by the measured
four-momenta of the particles in the final state~\cite{BESIII:2012rtd}.
The joint probability for observing $N$ events in the data samples is
\begin{equation}
    \mathcal{L} \equiv \prod_{i=1}^N \frac{\left|\sum_XA_X(\xi_i)\right|^2 \epsilon\left(\xi_i\right) \Phi\left(\xi_i\right)}{\sigma^{\prime}}
\end{equation}
where $A_X(\xi_i)$ is the amplitude corresponding to intermediate resonance X,
$\epsilon(\xi_i)$ is the detection efficiency,
$\Phi(\xi_i)$ is the standard element of phase space, $\sigma'$ is the normalization integral.

The fit procedure includes all possible intermediate states from the PDG~\cite{ParticleDataGroup:2022pth}
that satisfy $J^{PC}$ conservation in the subsequent two-body decay.
The likelihood contribution of background events is evaluated from the $\eta$ sideband region.
All possible amplitudes are tried in the fit, and only those
with statistical significance larger than 5$\sigma$ are retained.
The statistical significance of amplitude is estimated by incorporating the change
in log-likelihood as well as the degrees of freedom in the fits,
where the corresponding amplitude can be included or not.

The above strategy is implemented individually on the datasets at $\sqrt{s}=2.125$, 2.396,
and 2.900~GeV, which are the datasets with the highest signal yields.
The nominal solution is summarized in Table~\ref{tabsolution}.
The masses and widths of the $\rho(770)$ and $a_2(1320)$ in the PWA fit are determined by
scanning the likelihood value and are consistent with those in the PDG~\cite{ParticleDataGroup:2022pth}.
Since the contribution of other intermediate processes is relatively small,
their masses and widths are fixed to the PDG values.
The fit fractions of the intermediate states are also summarized in Table~\ref{tabsolution}.
The comparisons of invariant mass spectra and angular distributions between
data and the MC projections at $\sqrt{s}$=2.125, 2.396, and 2.900~GeV
are shown in Fig.2, Fig.3 and Fig.4 of the Supplemental Material~\cite{suppstuff}, respectively.
The MC results are consistent with data within statistical uncertainties.

\begin{table*}[!htb]
\begin{center}
\vspace*{-0.5cm}
\caption{\small Statistical significances and fit fractions of possible intermediate processes at $\sqrt{s}=2.125$, 2.396 and 2.900~GeV.}
\label{tabsolution}
\begin{tabular}{lcc|ccc|ccc} \hline
  \multicolumn{3}{c|}{$\sqrt{s}=2.125$~GeV}  &\multicolumn{3}{c|}{$\sqrt{s}=2.396$~GeV}  &\multicolumn{3}{c}{$\sqrt{s}=2.900$~GeV} \\ \hline
  Process         &Significance &Fraction (\%)   &Process            &Significance  &Fraction (\%)  &Process           &Significance  &Fraction (\%)   \\ \hline
  $\rho(770)\eta$ &$>20\sigma$  &$58.0\pm1.0$     &$\rho(770)\eta$    &$>20\sigma$   &$69.5\pm2.5$       &$\rho(770)\eta$    &$>20\sigma$   &$66.8\pm2.2$   \\
  $a_2(1320)\pi$  &$>20\sigma$  &$24.1\pm0.8$     &$a_2(1320)\pi$     &$>20\sigma$   &$13.0\pm1.1$       &$a_2(1320)\pi$     &$>10\sigma$   &$21.7\pm2.1$   \\
  $\rho(1450)\eta$&$>10\sigma$  &1.8$\pm0.3$      &$\rho(1450)\eta$   &$5.1\sigma$   &$1.0\pm0.4$        &$\rho(1450)\eta$   &$>10\sigma$   &$16.5\pm0.4$   \\
  $a_2(1700)\pi$  &$>10\sigma$  &$2.0\pm0.3$      &$\rho_3(1690)\eta$ &$9.7\sigma$   &$2.5\pm0.5$        &$\rho(1700)\eta$   &$6.5\sigma$   &$2.1\pm0.1$    \\
    -             & -           & -            &$a_2(1700)\pi$     &$6.8\sigma$   &$2.7\pm0.4$        & -                 & -            & -   \\
    -             & -           & -            &$\rho(1700)\eta$   &$5.8\sigma$   &$1.9\pm0.9$        & -                 & -            & -   \\
\hline
\end{tabular}
\end{center}
\end{table*}

Due to limited statistics, the above optimization strategy
is not performed for the other sixteen data samples.
Instead, the
same intermediate components are used as those found to be necessary at a nearby c.m. energy points with higher statistics.
The same intermediates as $\sqrt{s}=2.125$~GeV are used for the datasets
with $\sqrt{s}=2.000$, 2.050, 2.100, 2.150, 2.175, 2.200, and 2.232~GeV.
While the same intermediates as $\sqrt{s}=2.396$~GeV are used for datasets
with $\sqrt{s}=2.309$, 2.386, 2.644, and 2.646~GeV.
The same processes as 2.900~GeV are used in the fits to the remaining datasets.

The cross section, $\sigma$, at each c.m.\ energy is determined as
\begin{equation}
\label{eq:sigmaobs}
\sigma = \frac{N^{\rm obs}}{\mathcal{L}\cdot\epsilon \cdot {\mathcal B}\cdot (1+\delta^{\mathit{\gamma}})},
\end{equation}
where $N^{\rm{obs}}$ is the signal yield, $\mathcal{L}$ is the integrated luminosity of the dataset,
$\epsilon$ is the detection efficiency extracted from signal MC events which are generated
with the DIY generator derived from the PWA, ${\mathcal B}$ is the product of the
relevant daughter branching fractions, i.e.,
${\mathcal B}={\mathcal B}(\eta\to\gamma\gamma)=39.4\%$ for $\EPP$,
${\mathcal B}={\mathcal B}(\rho\to\pi\pi)\cdot{\mathcal B}(\eta\to\gamma\gamma)=39.4\%$
for $\RHOETA$ and
${\mathcal B}={\mathcal B}(a_2(1320)\to\eta\pi)\cdot{\mathcal B}(\eta\to\gamma\gamma)=5.7\%$
for $\API$,
$(1+\delta^{\mathit{\gamma}})$ is the ISR correction factor obtained from a QED calculation~\cite{Ping:2016pms,Kuraev:1985hb}
and incorporating the input cross sections in this analysis, where iterations are performed until the measured cross section converges. 

The signal yields for $\EPP$ are extracted from a simultaneous unbinned maximum-likelihood
fit to the $M(\gamma\gamma)$ spectra at each c.m.\ energy.
The signal is described by an MC-simulated shape convolved with a Gaussian function,
which is used to compensate for the differences in calibration and resolution between data and MC simulation.
The parameters of the Gaussian function are free.
A first-order Chebychev polynomial is used to describe the background.
Once the signal yields for $\EPP$ are obtained,
the signal yields for the intermediate processes of $\RHOETA$ and $\API$ are
determined by using their fractions derived from the PWA.

Below 2.3~GeV, the measured cross sections for $\EPP$ are systematically 30\% higher than those
of BaBar, as shown in Fig.~\ref{CS_etapipi}.
The resulting cross sections and related variables are listed in the Supplemental Material~\cite{suppstuff},
separately for $\EPP$ and its intermediate processes $\RHOETA$ and $\API$.

\begin{figure}[h]
	\centering
	\includegraphics[width=8.5cm]{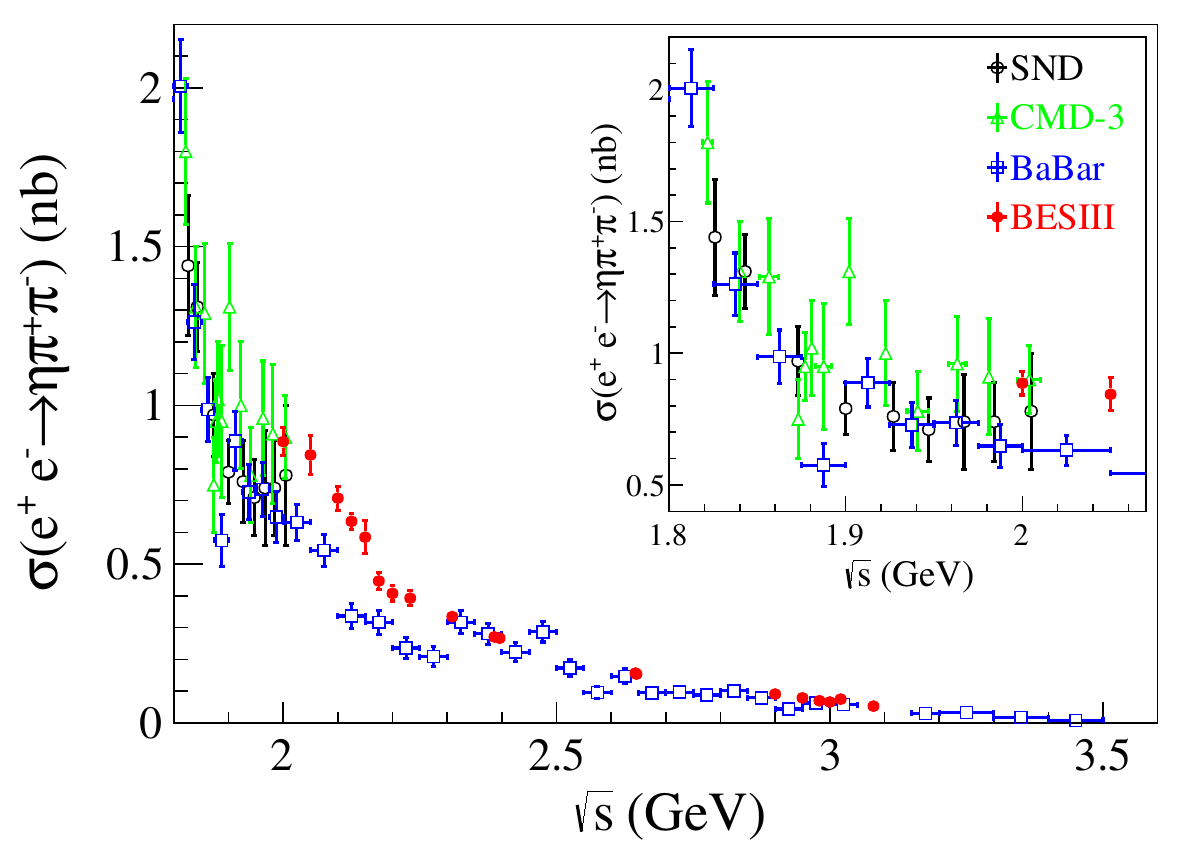}
    \vspace*{-0.3cm}
	\caption{ \small Cross sections of $\EPP$.
    Red solid dots with error bars are BESIII,
    blue hollow squares with error bars are BaBar~\cite{BaBar:2018erh},
    green hollow triangles with error bars are CMD-3~\cite{Gribanov:2019qgw},
    and black hollow dots with error bars are SND~\cite{Achasov:2017kqm}. The systematic uncertainties for BESIII are included.}
	\label{CS_etapipi}
\end{figure}

By implementing the same strategy described in Ref.~\cite{BESIII:2020vtu},
the following sources of systematic uncertainties on the measured cross sections are considered.
The uncertainty associated with the integrated luminosity, detection efficiency and PID for charged track,
and photon efficiency have been studied in Refs.\cite{BESIII:2017lkp,BESIII:2018ldc,BESIII:2018rdg}.
The helix parameters of the simulated charged tracks are corrected to match the resolution~\cite{BESIII:2012mpj},
and the difference with and without correction is taken as the systematic uncertainty related to the 4C kinematic fit.
The uncertainty regarding the ISR effect is obtained with
the accuracy of the radiation function~\cite{WorkingGrouponRadiativeCorrections:2010bjp},
and the contribution from the cross section lineshape, which is estimated by varying the
model parameters of the lineshape fit as performed in Refs.~\cite{BESIII:2018ldc,BESIII:2020vtu}.
The uncertainties originating from the $E/p$ ratio and $\eta$ helicity angle requirements
are estimated by the control samples of
$J/\psi\to\pi^+\pi^-\pi^0$ and $J/\psi\to K^+K^-\eta$, respectively.
The uncertainties related to the fit procedure are investigated by varying the fit range,
replacing the linear function for the background with a second-order polynomial function,
and varying the width of the Gaussian function for the signal.
The uncertainties from the branching fractions of intermediate states are taken from the PDG~\cite{ParticleDataGroup:2022pth}.
Uncertainty associated with the MC model for $\EPP$ cross sections is estimated by the alternative PWA model including all the components with significance of more than $3\sigma$.

The uncertainties due to the PWA fit mainly originate from the BW parametrization form, resonance parameters,  extra resonances, barrier factor and background estimation.
The uncertainties from the BW parametrization are estimated by replacing the
constant width in the
 relativistic BW function by a mass-dependent width.
The uncertainties related to the resonance parameters, which are taken from
the PDG and fixed in the nominal fit, are estimated by
changing the mass and width by one standard deviation of the PDG values.
The uncertainties associated with the extra resonances are estimated by
alternative fits including all the components with significance greater than $3\sigma$.
The uncertainties regarding the barrier factor~\cite{Chung:1993da,VonHippel:1972fg} are estimated by
varying the radius of the centrifugal barrier by one standard deviation,
assuming a uniform distribution of radius.
The uncertainties from the background estimation originate from the background shape and the background fraction, which are estimated by using the lower or upper sideband events and varying the fraction by $\pm\sigma$, respectively. The resulting largest differences to the nominal result are assigned as the systematic uncertainties.
The uncertainties from the PWA fit are strongly affected by the statistics.
Thus, those uncertainties of data with $\sqrt{s}=2.125$, $\sqrt{s}=2.396$,
and $\sqrt{s}=2.900$~GeV are assigned to their nearby c.m.\ energies.
Adding the systematic uncertainties in quadrature yields
the total systematic uncertainties of about 3.9\%-15.2\% for different data sets, which are summarized in Table IV for $\EPP$, Table V for $\RHOETA$ and Table VI for $\API$  in the Supplemental Material~\cite{suppstuff}.

To investigate the possible structures in the measured cross sections
of the quasi two-body process $\API$,
a $\chi^2$ fit incorporating the correlated and uncorrelated uncertainties among different c.m.\ energies is performed.
The fit probability density function is parameterized as the coherent sum of a continuum amplitude $f_1$
and a resonant amplitude $f_2$:
\begin{equation} \label{eq:fitfuntion}
\sigma(s) = \left|f_1 + e^{i\phi}f_2\right|^2,
\end{equation}
where $\phi$ is the relative phase angle between the amplitudes.
The amplitude $f_1$ is written as

\begin{equation} \label{eq:f1}
f_1 = C_0\cdot s^{-n}\sqrt{\Phi(\sqrt{s})},
\end{equation}
where $C_0\cdot s^{-n}$ describes the energy-dependent cross section of the continuum,
$\sqrt{\Phi(\sqrt{s})}$ is the two-body phase space, which takes the angular momentum and the width of the final state into account~\cite{Zou:2002ar}.
The resonant amplitude $f_2$ is described with a BW function as
\begin{equation} \label{eq:f2}
f_2 = \frac{\sqrt{12\pi\Gamma^{ee}_{R}\cdot\mathcal{B}_R\Gamma^{\rm tot}_{R}}}{s-M_{R}^2+iM_{R}\cdot\Gamma^{\rm tot}_{R}} \sqrt{\frac{\Phi(\sqrt{s})}{\Phi(M_{R})}} ,
\end{equation}
where $M_{R}$, $\Gamma_{R}^{ee}$ and $\Gamma^{\rm tot}_{R}$ are the mass,
partial width to $e^+e^-$ and total width of the assumed resonance $R$.
$\mathcal{B}_{R}$ is the branching fraction for $R \to a_2(1320)\pi$.

\begin{figure}[h!]
	\centering
	\begin{overpic}[width=8.5cm]{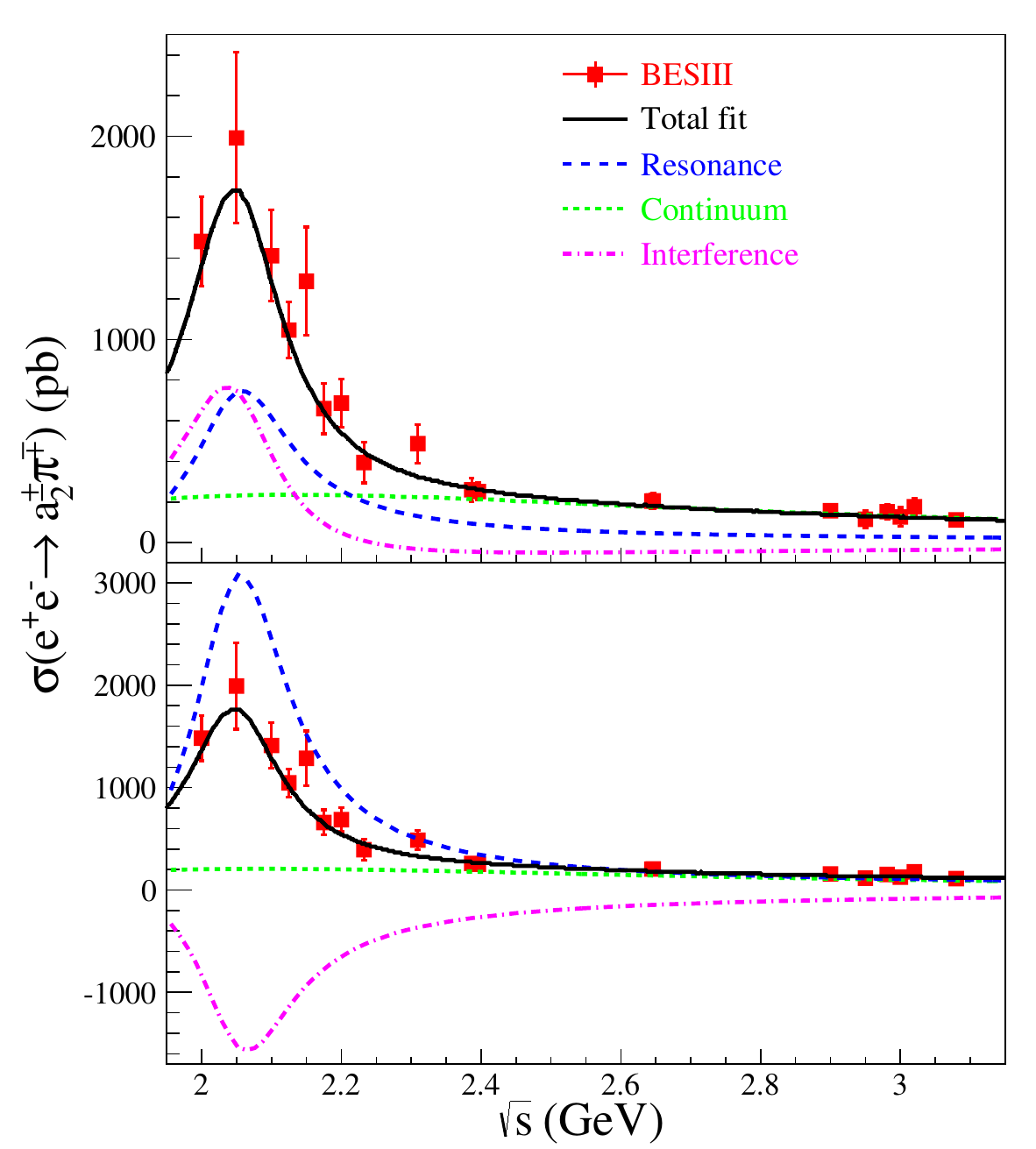}
	\put(35,85){(a)}
	\put(35,45){(b)}
	\end{overpic}
    \vspace*{-0.3cm}
	\caption{ \small Fits to the $\API$ cross sections.
	(a) Solution I, constructive interference. (b) Solution II, destructive interference.
Solid squares with error bars are BESIII data. The black solid curve is the total fit result, the blue dashed line
is the resonant component, the green dashed line is the continuum contribution,
and the magenta dot-dashed line represents the interference between the resonance and the continuum contribution.
The systematic uncertainties are included.}
	\label{fitCSa2pi}
\end{figure}

Fig.~\ref{fitCSa2pi} shows the results of the fits to the cross sections
for $\API$. There are two solutions with equal fit quality and
identical mass and width for the resonance, while the product
$\Gamma^{ee}_{R}\cdot\mathcal{B}_R$ and phase angle $\phi$ are different in the two solutions.
The goodness of fit is $\chi^{2}$/n.d.f. = 11.1/13 = 0.85,
where n.d.f.~is the number of degrees of freedom.
The fit parameters are summarized in Table~\ref{tabfita2pi}.
The statistical significance of this resonance is estimated to be
$5.5\sigma$ by comparing the change of $\chi^{2}$ ($\Delta \chi^{2}=40.4$)
with and without the $R$ amplitude in the fit
and taking the change of degrees of freedom ($\Delta$n.d.f.=4) into account.
\begin{table}[h!]
\begin{center}
\caption{Resonant parameters from the the fit to $\API$ cross sections.
The first uncertainty is statistical, and the second
one is systematic.}
\label{tabfita2pi}
\begin{tabular}{lcc} \hline
Parameter                                  & Solution 1    &Solution 2     \\  \hline
$M_{R}$ (MeV/$c^2$)                        & \multicolumn{2}{c}{$2044\pm31\pm4$}    \\
$\Gamma^{R}_{\rm{tot}}$ (MeV)              & \multicolumn{2}{c}{$163\pm69\pm24$}     \\
$\mathcal{B_{R}}\Gamma_{e^+e^-}^{R}$ (eV)  & $34.6\pm17.1\pm6.0$  &$137.1\pm73.3\pm2.1$ \\
$\phi$ (rad)                               & $1.95\pm0.97\pm0.06$ &$4.35\pm0.48\pm0.43$   \\ \hline
\end{tabular}
\end{center}
\end{table}
\begin{figure}[h!]
	\centering
	\begin{overpic}[width=8.5cm]{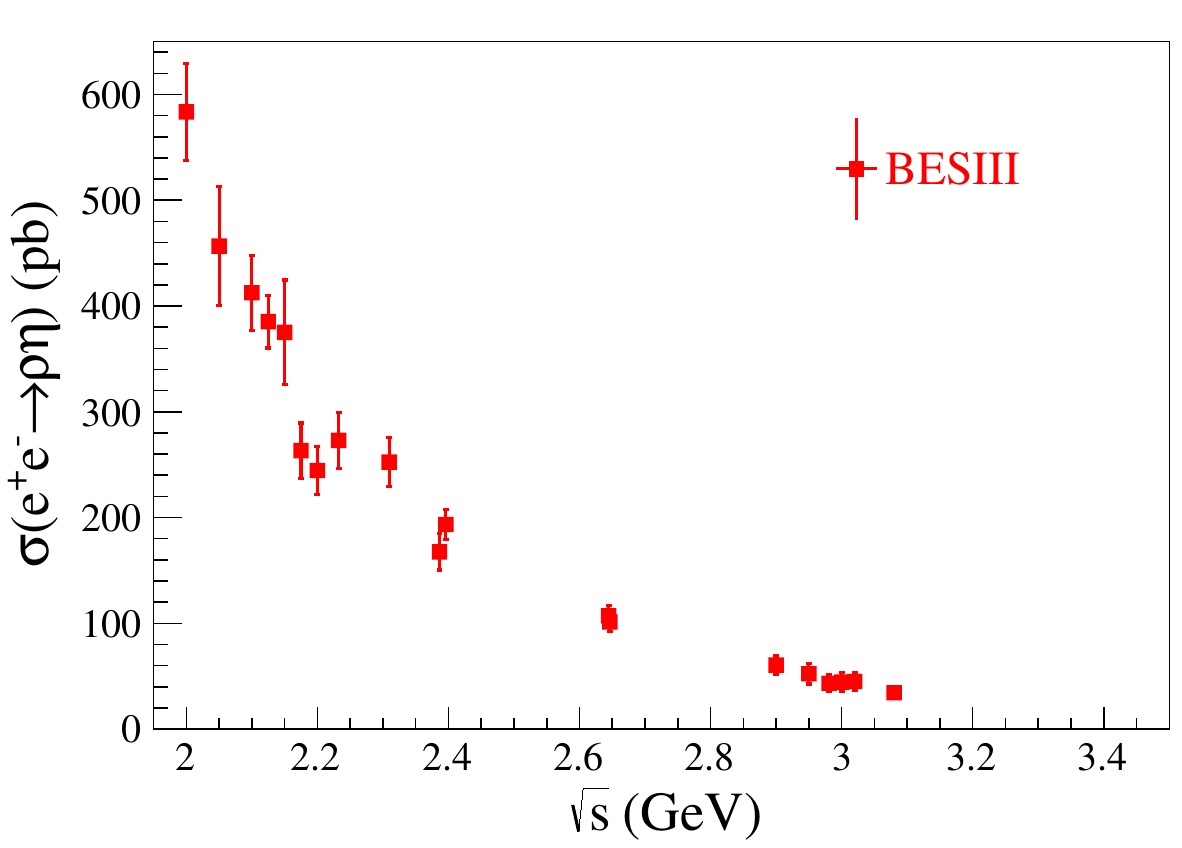}
	\end{overpic}
    \vspace*{-0.5cm}
	\caption{ \small Cross sections of $\RHOETA$.  Solid squares with error bars are BESIII data.}
	\label{fitCSrhoeta}
\end{figure}
The uncertainty of the parametrization of the continuum contribution for $\API$ is estimated
by replacing the $s-$dependent continuum function $C_{1}/s^{n}$ with an exponential function
of the form ${C_{1}\cdot}e^{-n(\sqrt{s}-M_{\rm{th}})}$, where $M_{\rm{th}}=m_{a_2}+m_{\pi}$.
The differences of the obtained mass and width, which are 4~MeV/$c^{2}$ and 24~MeV, respectively, are taken as the systematic uncertainties.

Fig.~\ref{fitCSrhoeta} shows the cross sections for $\RHOETA$.
A dip structure around 2180~MeV/$c^2$ is observed with statistical significance of $3.0\sigma$.
Unfortunately, exploratory studies are not able to extract a
physical resonance state. To make any conclusive statement,
a finer energy scan is needed in this particular region.

In summary, we present a PWA of the process $\EPP$ using data samples
collected by the BESIII detector at 19 c.m.\ energies between 2.00 and 3.08~GeV.
The cross sections for the process $\EPP$,
and its subprocesses $\RHOETA$ and $\API$ have been measured.
The obtained cross sections of $\EPP$ are systematically higher than those of BaBar~\cite{BaBar:2018erh} by more than 30\%
at $\sqrt{s}$ between $2.00$ and 2.30~GeV.
A coherent fit to the cross section lineshape for the process $\API$ is performed using a resonant amplitude and a continuum amplitude. One resonant structure is observed with a significance of $5.5\sigma$.
Its mass, width and $\mathcal{B_{R}}\cdot\Gamma_{e^+e^-}^{R}$ are
determined to be ($2044\pm31\pm4$)~MeV/$c^2$, ($163\pm69\pm24$)~MeV and ($34.6\pm17.1\pm6.0$) or ($137.1\pm73.3\pm2.1$)~eV, respectively. The observed structure agrees with the properties of the $\rho(2000)$ resonance observed in $e^+e^-\to\omega\pi^0$~\cite{BESIII:2020xmw}, which indicates the first observation of the decay $\rho(2000)\to a_{2}(1320)\pi$.
To further understand the dip structure around 2180~MeV/$c^2$ observed in the $\RHOETA$ cross section lineshape, it will be necessary to acquire more energy points in order to improve the precision of the cross section measurements in the future.

The BESIII Collaboration thanks the staff of BEPCII and the IHEP computing center for their strong support.
This work is supported in part by National Key R\&D Program of China under Contracts
No. 2020YFA0406400, No. 2020YFA0406300; National Natural Science Foundation of China (NSFC) under Contracts Nos. 11635010, 11735014, 11835012, 11935015, 11935016, 11935018, 11961141012, 12025502, 12035009, 12035013, 12061131003, 12192260, 12192261, 12192262, 12192263, 12192264, 12192265, 12221005, 12225509, 12235017, 12205255, 11975118; the Chinese Academy of Sciences (CAS) Large-Scale Scientific Facility Program; the CAS Center for Excellence in Particle Physics (CCEPP); Joint Large-Scale Scientific Facility Funds of the NSFC and CAS under Contract No. U1832207, No. U1732263, No. U2032105; CAS Key Research Program of Frontier Sciences under Contracts Nos. QYZDJ-SSW-SLH003, QYZDJ-SSW-SLH040; 100 Talents Program of CAS; Natural Science Foundation of Gansu Province under Contract No. 23JRRA578;
The Institute of Nuclear and Particle Physics (INPAC) and Shanghai Key Laboratory for Particle Physics and Cosmology; European Union's Horizon 2020 research and innovation programme under Marie Sklodowska-Curie grant agreement under Contract No. 894790; German Research Foundation DFG under Contracts Nos. 455635585, Collaborative Research Center CRC 1044, FOR5327, GRK 2149; Istituto Nazionale di Fisica Nucleare, Italy; Ministry of Development of Turkey under Contract No. DPT2006K-120470; National Research Foundation of Korea under Contract No. NRF-2022R1A2C1092335; National Science and Technology fund of Mongolia; National Science Research and Innovation Fund (NSRF) via the Program Management Unit for Human Resources \& Institutional Development, Research and Innovation of Thailand under Contract No. B16F640076; Polish National Science Centre under Contract No. 2019/35/O/ST2/02907; The Swedish Research Council; U. S. Department of Energy under Contract No. DE-FG02-05ER41374.

\bibliography{reference}

\end{document}


\newcommand{\ks}{K_{S}^{0}}
\newcommand{\EP}{e^{+}}
\newcommand{\EM}{e^{-}}
\newcommand{\epm}{e^{\pm}}
\newcommand{\vpho}{\gamma^{\ast}}
\newcommand{\qqbar}{q\bar{q}}

\newcommand{\ee}{e^{+}e^{-}}
\newcommand{\mm}{\mu^{+}\mu^{-}}
\newcommand{\alfs}{\alpha_{s}}
\newcommand{\alfmz}{\alpha(M_{Z}^{2})}
\newcommand{\amu}{a_{\mu}}
\newcommand{\Lam}{\Lambda_{c}}
\newcommand{\lam}{\Lambda_{c}^{+}}
\newcommand{\lambar}{\bar{\Lambda}_{c}^{-}}
\newcommand{\Lambdac}{\Lambda_{c}}
\newcommand{\mbc}{M_{BC}}
\newcommand{\dele}{\Delta E}
\newcommand{\ebm}{E_{\textmd{beam}}}
\newcommand{\ecm}{E_{\textmd{c.m.}}}
\newcommand{\pbm}{p_{\textmd{beam}}}
\newcommand{\MuMu}{\mu\mu}
\newcommand{\mumu}{\mu\mu}
\newcommand{\tata}{\tau^{+}\tau^{-}}
\newcommand{\pipi}{\pi^{+}\pi^{-}}
\newcommand{\gaga}{\gamma\gamma}
\newcommand{\twopho}{\ee+X}
\newcommand{\sqs}{\sqrt{s}}
\newcommand{\sqsp}{\sqrt{s^{\prime}}}
\newcommand{\da}{\Delta\alpha}
\newcommand{\das}{\Delta\alpha(s)}
\newcommand{\dimu}{\ee \ra \mumu}
\newcommand{\dedx}{\textmd{d}E/\textmd{d}x}
\newcommand{\chip}{\chi_{\textmd{Prob}}}
\newcommand{\chiP}{\chi_{p}}
\newcommand{\evz}{V_{z}^{\textmd{evt}}}
\newcommand{\evzloose}{V_{z,\textmd{loose}}^{\textmd{evt}}}
\newcommand{\avz}{V_{z}^{\textmd{ave}}}
\newcommand{\Ngd}{N_{\textmd{good}}}
\newcommand{\Ncru}{N_{\textmd{crude}}}
\newcommand{\pio}{\pi^{0}}
\newcommand{\rpid}{r_{\textmd{PID}}}

\newcommand{\Nhxobs}{N_{h+X}^{\textmd{obs}}}
\newcommand{\Nhobs}{N_{h}^{\textmd{obs}}}
\newcommand{\Npioxobs}{N_{\pi^{0}+X}^{\textmd{obs}}}
\newcommand{\Nksxobs}{N_{\ks+X}^{\textmd{obs}}}
\newcommand{\Npioobs}{N_{\pi^{0}}^{\textmd{obs}}}
\newcommand{\Nksobs}{N_{\ks}^{\textmd{obs}}}
\newcommand{\Nhxtru}{N_{h+X}^{\textmd{tru}}}
\newcommand{\Nhtru}{N_{h}^{\textmd{tru}}}
\newcommand{\Npiotru}{N_{\pi^{0}}^{\textmd{tru}}}
\newcommand{\Nkstru}{N_{\ks}^{\textmd{tru}}}
\newcommand{\Nbarhxobs}{\bar{N}_{h+X}^{\textmd{obs}}}
\newcommand{\Nbarhobs}{\bar{N}_{h}^{\textmd{obs}}}
\newcommand{\Nbarpioobs}{\bar{N}_{\pi^{0}}^{\textmd{obs}}}
\newcommand{\Nbarksobs}{\bar{N}_{\ks}^{\textmd{obs}}}
\newcommand{\Nbarhxtru}{\bar{N}_{h+X}^{\textmd{tru}}}
\newcommand{\Nbarhtru}{\bar{N}_{h}^{\textmd{tru}}}
\newcommand{\Nbarpiotru}{\bar{N}_{\pi^{0}}^{\textmd{tru}}}
\newcommand{\Nbarkstru}{\bar{N}_{\ks}^{\textmd{tru}}}

\newcommand{\Nhadtot}{N_{\textmd{had}}^{\textmd{tot}}}
\newcommand{\Nhadobs}{N_{\textmd{had}}^{\textmd{obs}}}
\newcommand{\Nbarhadobs}{\bar{N}_{\textmd{had}}^{\textmd{obs}}}
\newcommand{\Nhadtru}{N_{\textmd{had}}^{\textmd{tru}}}
\newcommand{\Nbarhadtru}{\bar{N}_{\textmd{had}}^{\textmd{tru}}}
\newcommand{\Nhadphy}{N_{\textmd{had}}}

\newcommand{\cshadobs}{\sigma_{\textmd{had}}^{\textmd{obs}}}
\newcommand{\effhad}{\vap_{\textmd{had}}}
\newcommand{\efftrg}{\vap_{\textmd{trig}}}
\newcommand{\lint}{\mathcal{L}_{\textmd{int}}}
\newcommand{\Nbkg}{N_{\textmd{bkg}}}
\newcommand{\NbkgTot}{N_{\textrm{bkg}}^{\textrm{Tot}}}
\newcommand{\csbkg}{\sigma_{\textmd{bkg}}}
\newcommand{\Nmcsur}{N_{\textmd{MC}}^{\textmd{sur}}}
\newcommand{\Nmcsurori}{N_{\textmd{MC}}^{\textmd{sur,nom.}}}
\newcommand{\Nmcsurwtd}{N_{\textmd{MC}}^{\textmd{sur,wtd.}}}
\newcommand{\Nmcgen}{N_{\textmd{MC}}^{\textmd{gen}}}
\newcommand{\vap}{\varepsilon}
\newcommand{\chisq}{\chi^{2}}
\newcommand{\cshadphy}{\sigma_{\textmd{had}}^{\textmd{phy}}}
\newcommand{\cshadtot}{\sigma_{\textmd{had}}^{\textmd{tot}}}
\newcommand{\cshadborn}{\sigma_{\textmd{had}}^{0}}
\newcommand{\cshadborncon}{\sigma_{\textmd{con}}^{0}}
\newcommand{\cshadbornres}{\sigma_{\textmd{res}}^{0}}
\newcommand{\csdimuborn}{\sigma_{\mu\mu}^{0}}
\newcommand{\rpqcd}{R_{\textmd{pQCD}}}
\newcommand{\Nprod}{N_{\textmd{prod}}}
\newcommand{\Nhadnet}{N_{\textrm{had}}^{\textrm{net}}}
\newcommand{\Delrel}{\Delta_{\textrm{rel}}}

\newcommand{\fourpionchg}{\pipi\pipi}
\newcommand{\fourpionneu}{\pipi\pi^{0}\pi^{0}}
\newcommand{\sixpionchg}{3(\pipi)}
\newcommand{\thrpionneu}{\pipi\pi^{0}}
\newcommand{\twopionchg}{\pipi}

\newcommand{\Nsurnpion}{N_{\textmd{sur}}^{n\pi}}
\newcommand{\Ngennpion}{N_{\textmd{gen}}^{n\pi}}
\newcommand{\Ngentot}{N_{\textmd{gen}}^{\textmd{tot}}}
\newcommand{\effincnpion}{\vap_{n\pi}^{\textmd{inc}}}
\newcommand{\effincnpionp}{\vap_{n\pi}^{\textmd{inc},\prime}}
\newcommand{\effincnonnpion}{\vap_{\textmd{non}-n\pi}^{\textmd{inc}}}
\newcommand{\effexcnpion}{\vap_{n\pi}^{\textmd{exc}}}
\newcommand{\fracnpion}{f_{n\pi}}
\newcommand{\fracnpionp}{f_{n\pi}^{\prime}}
\newcommand{\fracnonnpion}{f_{\textmd{non}-n\pi}}

\newcommand{\Nsurtwopion}{N_{\textmd{sur}}^{2\pi}}
\newcommand{\Ngentwopion}{N_{\textmd{gen}}^{2\pi}}
\newcommand{\effinctwopion}{\vap_{2\pi}^{\textmd{inc}}}
\newcommand{\effinctwopionp}{\vap_{2\pi}^{\textmd{inc},\prime}}
\newcommand{\effincnontwopion}{\vap_{\textmd{non}-2\pi}^{\textmd{inc}}}
\newcommand{\effexctwopion}{\vap_{2\pi}^{\textmd{exc}}}
\newcommand{\fractwopion}{f_{2\pi}}
\newcommand{\fractwopionp}{f_{2\pi}^{\prime}}
\newcommand{\fracnontwopion}{f_{\textmd{non}-2\pi}}

\newcommand{\Nsurthrpion}{N_{\textmd{sur}}^{3\pi}}
\newcommand{\Ngenthrpion}{N_{\textmd{gen}}^{3\pi}}
\newcommand{\effincthrpion}{\vap_{3\pi}^{\textmd{inc}}}
\newcommand{\effincthrpionp}{\vap_{3\pi}^{\textmd{inc},\prime}}
\newcommand{\effincnonthrpion}{\vap_{\textmd{non}-3\pi}^{\textmd{inc}}}
\newcommand{\effexcthrpion}{\vap_{3\pi}^{\textmd{exc}}}
\newcommand{\fracthrpion}{f_{3\pi}}
\newcommand{\fracthrpionp}{f_{3\pi}^{\prime}}
\newcommand{\fracnonthrpion}{f_{\textmd{non}-3\pi}}

\newcommand{\Nsurfourpion}{N_{\textmd{sur}}^{4\pi}}
\newcommand{\Ngenfourpion}{N_{\textmd{gen}}^{4\pi}}
\newcommand{\effincfourpion}{\vap_{4\pi}^{\textmd{inc}}}
\newcommand{\effincfourpionp}{\vap_{4\pi}^{\textmd{inc},\prime}}
\newcommand{\effincnonfourpion}{\vap_{\textmd{non}-4\pi}^{\textmd{inc}}}
\newcommand{\effexcfourpion}{\vap_{4\pi}^{\textmd{exc}}}
\newcommand{\fracfourpion}{f_{4\pi}}
\newcommand{\fracfourpionp}{f_{4\pi}^{\prime}}
\newcommand{\fracnonfourpion}{f_{\textmd{non}-4\pi}}

\newcommand{\Npionprod}{N_{\textmd{prod}}^{4\pi}}
\newcommand{\Ndatasur}{N_{\textmd{data}}^{\textmd{sur}}}
\newcommand{\Nobspion}{N_{\textmd{obs}}^{4\pi}}
\newcommand{\Nhadprod}{N_{\textmd{prod}}^{\textmd{had}}}
\newcommand{\sigmaobs}{\sigma_{\textmd{obs}}}
\newcommand{\effhadp}{\vap_{\textmd{had}}^{\prime}}

\newcommand{\effpion}{\vap_{4\pi}}
\newcommand{\effexcpion}{\vap_{4\pi}^{\textmd{exc}}}
\newcommand{\effincpion}{\vap_{4\pi}^{\textmd{inc}}}
\newcommand{\effincpionI}{\vap_{4\pi}^{\textmd{inc},1}}
\newcommand{\effincpionII}{\vap_{4\pi}^{\textmd{inc},2}}
\newcommand{\effincpionp}{\vap_{4\pi}^{\textmd{inc},\prime}}
\newcommand{\effincremain}{\vap_{\textmd{non}-n\pi}^{\textmd{inc}}}

\newcommand{\fracpion}{f_{4\pi}}
\newcommand{\fracnonpion}{f_{\textmd{non}-4\pi}}
\newcommand{\fracnonpionp}{f_{\textmd{non}-4\pi}^{\prime}}
\newcommand{\fracpionII}{f_{4\pi}^{2}}
\newcommand{\fracpionp}{f_{4\pi}^{\prime}}
\newcommand{\reladiff}{\Delta_{\textmd{rel}}}

\newcommand{\Nsursixpion}{N_{\textmd{sur}}^{6\pi}}
\newcommand{\Ngensixpion}{N_{\textmd{gen}}^{6\pi}}
\newcommand{\effincsixpion}{\vap_{6\pi}^{\textmd{inc}}}
\newcommand{\fracsixpion}{f_{6\pi}}

\newcommand{\etot}{E_{\textmd{tot}}}
\newcommand{\ptot}{p_{\textmd{tot}}}
\newcommand{\plab}{p_{\textmd{Lab}}}
\newcommand{\mpiOI}{M(\pi^{0}_{1})}
\newcommand{\mpiOII}{M(\pi^{0}_{2})}

\newcommand{\widtheeoi}{\varGamma^{\textmd{ee}}_{0,i}}
\newcommand{\widtheeoj}{\varGamma^{\textmd{ee}}_{0,j}}
\newcommand{\widtheeo}{\varGamma^{\textmd{ee}}_{0}}
\newcommand{\widthee}{\varGamma^{\textmd{ee}}}
\newcommand{\widtheeexpi}{\varGamma^{\textmd{ee}}_{\textmd{exp},i}}
\newcommand{\widtheeexp}{\varGamma^{\textmd{ee}}_{\textmd{exp}}}
\newcommand{\widthtoti}{\varGamma^{\textmd{tot}}_{i}}
\newcommand{\widthtot}{\varGamma^{\textmd{tot}}}

\newcommand{\vpqed}{\Pi_{\textmd{QED}}}
\newcommand{\vpqcd}{\Pi_{\textmd{QCD}}}
\newcommand{\vpcon}{\Pi_{\textmd{con}}}
\newcommand{\vpres}{\Pi_{\textmd{res}}}
\newcommand{\vpo}{\Pi_{0}}
\newcommand{\rcon}{R_{\textmd{con}}}
\newcommand{\rres}{R_{\textmd{res}}}
\newcommand{\rexp}{R_{\textmd{exp}}}

\newcommand{\delvert}{\delta_{\textmd{vert}}}
\newcommand{\delvp}{\delta_{\textmd{vac}}}
\newcommand{\delbrem}{\delta_{\gamma}}
\newcommand{\delobs}{\delta_{\textmd{obs}}}
\newcommand{\radiatorsf}{F_{\textmd{SF}}}
\newcommand{\radiatorfd}{F_{\textmd{FD}}}
\newcommand{\DelFD}{\Delta_{\textmd{FD}}}
\newcommand{\DelFDCal}{\Delta_{\textmd{cal}}}
\newcommand{\DelFDcs}{\Delta_{\sigma}}
\newcommand{\DelFDvp}{\Delta_{\textmd{vp}}}

\newcommand{\costh}{\cos\theta}
\newcommand{\costhIprg}{\cos\theta_{\textmd{1prg}}}
\newcommand{\costhIIprg}{\cos\theta_{\textmd{2prg}}}
\newcommand{\costhIIIprg}{\cos\theta_{\textmd{3prg}}}
\newcommand{\costhIVprg}{\cos\theta_{\textmd{4prg}}}
\newcommand{\costhrestprg}{\cos\theta_{\textmd{restprg}}}
\newcommand{\emce}{E^{\textmd{ctrk.}}_{\textmd{emc}}}
\newcommand{\emceIprg}{E^{\textmd{ctrk.}}_{\textmd{emc,1prg}}}
\newcommand{\emceIIprg}{E^{\textmd{ctrk.}}_{\textmd{emc,2prg}}}
\newcommand{\emceIIIprg}{E^{\textmd{ctrk.}}_{\textmd{emc,3prg}}}
\newcommand{\emceIVprg}{E^{\textmd{ctrk.}}_{\textmd{emc,4prg}}}
\newcommand{\emcerestprg}{E^{\textmd{ctrk.}}_{\textmd{emc,restprg}}}
\newcommand{\isocosth}{\cos\theta_{\textmd{iso}}}
\newcommand{\isocosthIprg}{\cos\theta_{\textmd{iso,1prg}}}
\newcommand{\isocosthIIprg}{\cos\theta_{\textmd{iso,2prg}}}
\newcommand{\isocosthIIIprg}{\cos\theta_{\textmd{iso,3prg}}}
\newcommand{\isocosthIVprg}{\cos\theta_{\textmd{iso,4prg}}}
\newcommand{\isocosthrestprg}{\cos\theta_{\textmd{iso,restprg}}}
\newcommand{\eop}{E/P}
\newcommand{\eopIprg}{E/P_{\textmd{1prg}}}
\newcommand{\eopIIprg}{E/P_{\textmd{2prg}}}
\newcommand{\eopIIIprg}{E/P_{\textmd{3prg}}}
\newcommand{\eopIVprg}{E/P_{\textmd{4prg}}}
\newcommand{\eoprestprg}{E/P_{\textmd{restprg}}}
\newcommand{\nisogam}{N_{\textmd{isogam}}}
\newcommand{\nisogamIprg}{N_{\textmd{isogam,1prg}}}
\newcommand{\nisogamIIprg}{N_{\textmd{isogam,2prg}}}
\newcommand{\nisogamIIIprg}{N_{\textmd{isogam,3prg}}}
\newcommand{\nisogamIVprg}{N_{\textmd{isogam,4prg}}}
\newcommand{\nisogamrestprg}{N_{\textmd{isogam,restprg}}}
\newcommand{\ptrk}{p_{\textmd{ctrk}}}
\newcommand{\pIprg}{p^{\textmd{ctrk}}_{\textmd{1prg}}}
\newcommand{\pIIprg}{p^{\textmd{ctrk}}_{\textmd{2prg}}}
\newcommand{\pIIIprg}{p^{\textmd{ctrk}}_{\textmd{3prg}}}
\newcommand{\pIVprg}{p^{\textmd{ctrk}}_{\textmd{4prg}}}
\newcommand{\prestprg}{p^{\textmd{ctrk}}_{\textmd{restprg}}}
\newcommand{\tote}{E_{\textmd{vis.}}}
\newcommand{\totevte}{E_{\textmd{tot.}}}
\newcommand{\totevteIprg}{E_{\textmd{tot.}}^{\textmd{1prg}}}
\newcommand{\balanceIprg}{\textmd{Balance}}
\newcommand{\ngamma}{N_{\gamma}}
\newcommand{\ngammaIprg}{N_{\gamma,\textmd{1prg}}}
\newcommand{\ngammaIIprg}{N_{\gamma,\textmd{2prg}}}
\newcommand{\ngammaIIIprg}{N_{\gamma,\textmd{3prg}}}
\newcommand{\ngammaIVprg}{N_{\gamma,\textmd{4prg}}}
\newcommand{\ngammarestprg}{N_{\gamma,\textmd{restprg}}}
\newcommand{\ngoodwt}{N_{\textmd{good}}^{\textmd{Wt}}}
\newcommand{\ngood}{N_{\textmd{good}}}
\newcommand{\npiO}{N_{\pi^{0}}}
\newcommand{\npiOIprg}{N_{\pi^{0}}^{\textmd{1prg}}}
\newcommand{\npiOIIprg}{N_{\pi^{0}}^{\textmd{2prg}}}
\newcommand{\npp}{N_{p}}
\newcommand{\npm}{N_{\bar{p}}}
\newcommand{\nkp}{N_{K^{+}}}
\newcommand{\nkm}{N_{K^{-}}}
\newcommand{\npip}{N_{\pi^{+}}}
\newcommand{\npim}{N_{\pi^{-}}}
\newcommand{\ppp}{P(p^{+})}
\newcommand{\ppm}{P(\bar{p}^{-})}
\newcommand{\pkp}{p(K^{+})}
\newcommand{\pkm}{p(K^{-})}
\newcommand{\ppip}{P(\pi^{+})}
\newcommand{\ppim}{P(\pi^{-})}
\newcommand{\ppiO}{P(\pi^{0})}
\newcommand{\mpiO}{M(\pi^{0})}
\newcommand{\mks}{M(K^{0}_{s})}
\newcommand{\pks}{p_{K^{0}_{s}}}
\newcommand{\mphi}{M(\phi)}
\newcommand{\pphi}{p_{\phi}}
\newcommand{\mIIgam}{M(\gamma\gamma)}
\newcommand{\mIIgamIprg}{M(\gamma\gamma)^{\textmd{1prg}}}
\newcommand{\pIIgam}{p_{\gamma\gamma}}
\newcommand{\mlambda}{M(\Lambda)}
\newcommand{\plambda}{p_{\Lambda}}
\newcommand{\mdO}{M(D^{0})}
\newcommand{\pdO}{p_{D^{0}}}
\newcommand{\mdstarO}{M(D^{\ast 0})}
\newcommand{\pdstarO}{p_{D^{\ast 0}}}
\newcommand{\mdp}{M(D^{\pm})}
\newcommand{\pdp}{p_{D^{\pm}}}
\newcommand{\mdstarp}{M(D^{\ast\pm})}
\newcommand{\pdstarp}{p_{D^{\ast\pm}}}
\newcommand{\mds}{M(D_{s}^{\pm})}
\newcommand{\pds}{p_{D_{s}^{\pm}}}
\newcommand{\mdstars}{M(D_{s}^{\ast\pm})}
\newcommand{\pdstars}{p_{D_{s}^{\ast\pm}}}
\newcommand{\Vr}{V_{r}}
\newcommand{\Vz}{V_{z}}

\newcommand{\gev}{\mathrm{GeV}}
\newcommand{\mev}{\mathrm{MeV}}
\newcommand{\mevcc}{\mathrm{MeV}/c^{2}}
\newcommand{\gevc}{\mathrm{GeV}/c}
\newcommand{\gevcc}{\mathrm{GeV}/c^2}

\newcommand{\nchg}{N_{\textmd{chg}}}
\newcommand{\eff}{\vap}

\newcommand{\critecm}{1.780}

\newcommand{\ENERGYAT}{4575.5}
\newcommand{\ENERGYBT}{4575.5}
\newcommand{\ENERGYCT}{4575.5}
\newcommand{\ENERGYDT}{4575.5}
\newcommand{\ksdecay}{\ks\ra\pi^{+}\pi^{-}}
\newcommand{\phidecay}{\phi\ra K^{+}K^{-}}
\newcommand{\piOdecay}{\pi^{0}\ra\gamma\gamma}
\newcommand{\Lambdadecay}{\Lambda\ra p\pi^{-}}
\newcommand{\DOdecay}{D^{0}\ra K^{-}\pi^{+}}
\newcommand{\DStarOdecay}{D^{\ast0}\ra D^{0}\pi^{0}}
\newcommand{\Dpdecay}{D^{+}\ra K^{+}\pi^{+}\pi^{-}}
\newcommand{\DStarpdecay}{D^{\ast+}\ra D^{0}\pi^{+}}
\newcommand{\Dsdecay}{D^{+}_{s}\ra K^{+}K^{-}\pi^{+}}
\newcommand{\DStarsdecay}{D^{\ast+}_{s}\ra D^{+}_{s}\gamma}


\title{\boldmath \textbf{Measurement of the cross sections for $e^+e^-\to\eta\pi^+\pi^-$ at center-of-mass energies between 2.00 and 3.08~GeV} }

\author{
M.~Ablikim$^{1}$, M.~N.~Achasov$^{4,b}$, P.~Adlarson$^{75}$, X.~C.~Ai$^{80}$, R.~Aliberti$^{35}$, A.~Amoroso$^{74A,74C}$, M.~R.~An$^{39}$, Q.~An$^{71,58}$, Y.~Bai$^{57}$, O.~Bakina$^{36}$
% ...
\\
\vspace{0.2cm}
(BESIII Collaboration)\\
\vspace{0.2cm} {\it
$^{1}$ Institute of High Energy Physics, Beijing 100049, People's Republic of China\\
% ...
$^{4}$ Budker Institute of Nuclear Physics SB RAS (BINP), Novosibirsk 630090, Russia\\
% ...
$^{35}$ Johannes Gutenberg University of Mainz, Johann-Joachim-Becher-Weg 45, D-55099 Mainz, Germany\\
$^{36}$ Joint Institute for Nuclear Research, 141980 Dubna, Moscow region, Russia\\
% ...
$^{39}$ Liaoning Normal University, Dalian 116029, People's Republic of China\\
% ...
$^{57}$ Southeast University, Nanjing 211100, People's Republic of China\\
$^{58}$ State Key Laboratory of Particle Detection and Electronics, Beijing 100049, Hefei 230026, People's Republic of China\\
% ...
$^{71}$ University of Science and Technology of China, Hefei 230026, People's Republic of China\\
% ...
$^{74}$ University of Turin and INFN, (A)University of Turin, I-10125, Turin, Italy; (B)University of Eastern Piedmont, I-15121, Alessandria, Italy; (C)INFN, I-10125, Turin, Italy\\
$^{75}$ Uppsala University, Box 516, SE-75120 Uppsala, Sweden\\
% ...
$^{80}$ Zhengzhou University, Zhengzhou 450001, People's Republic of China\\
\vspace{0.2cm}
% ...
$^{b}$ Also at the Novosibirsk State University, Novosibirsk, 630090, Russia\\
% ...
}
}

\maketitle{}


\begin{figure}[!htb]
	\centering
        \includegraphics[width=7.5cm]{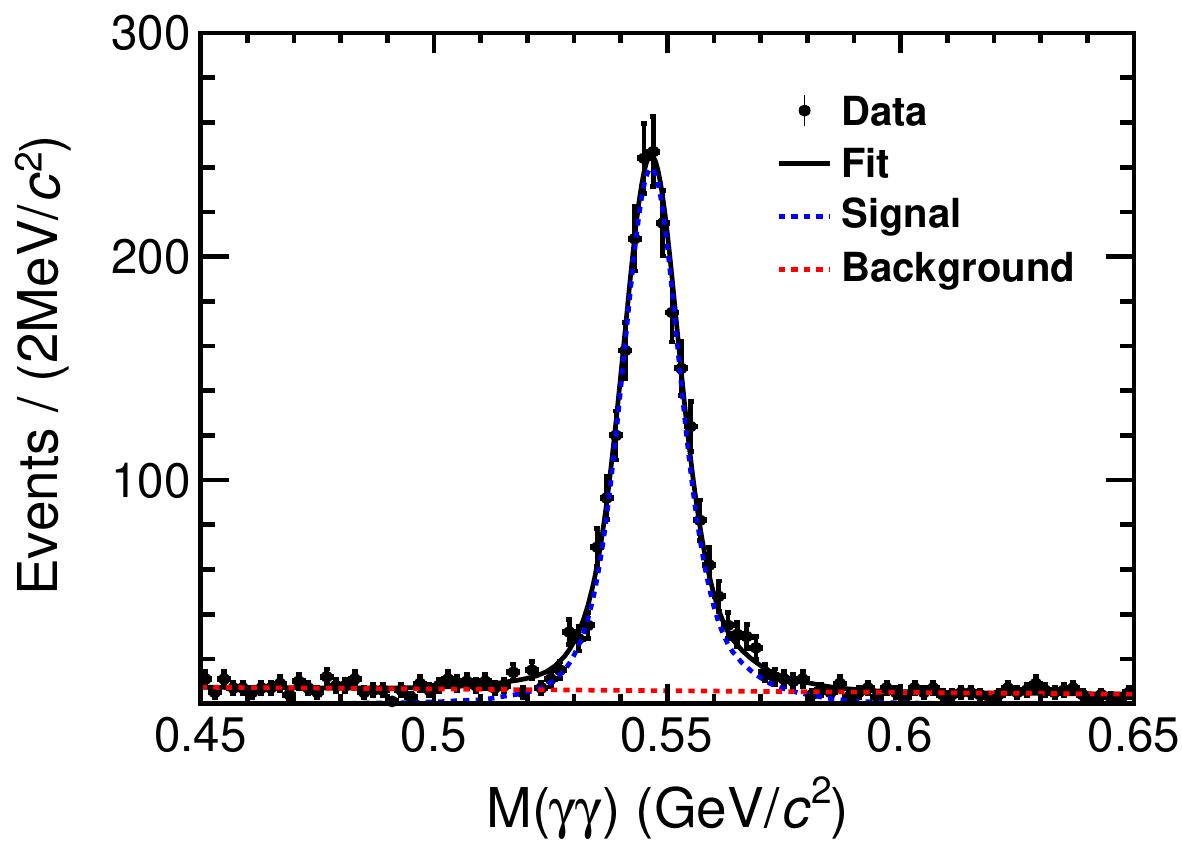}
        \includegraphics[width=7.5cm]{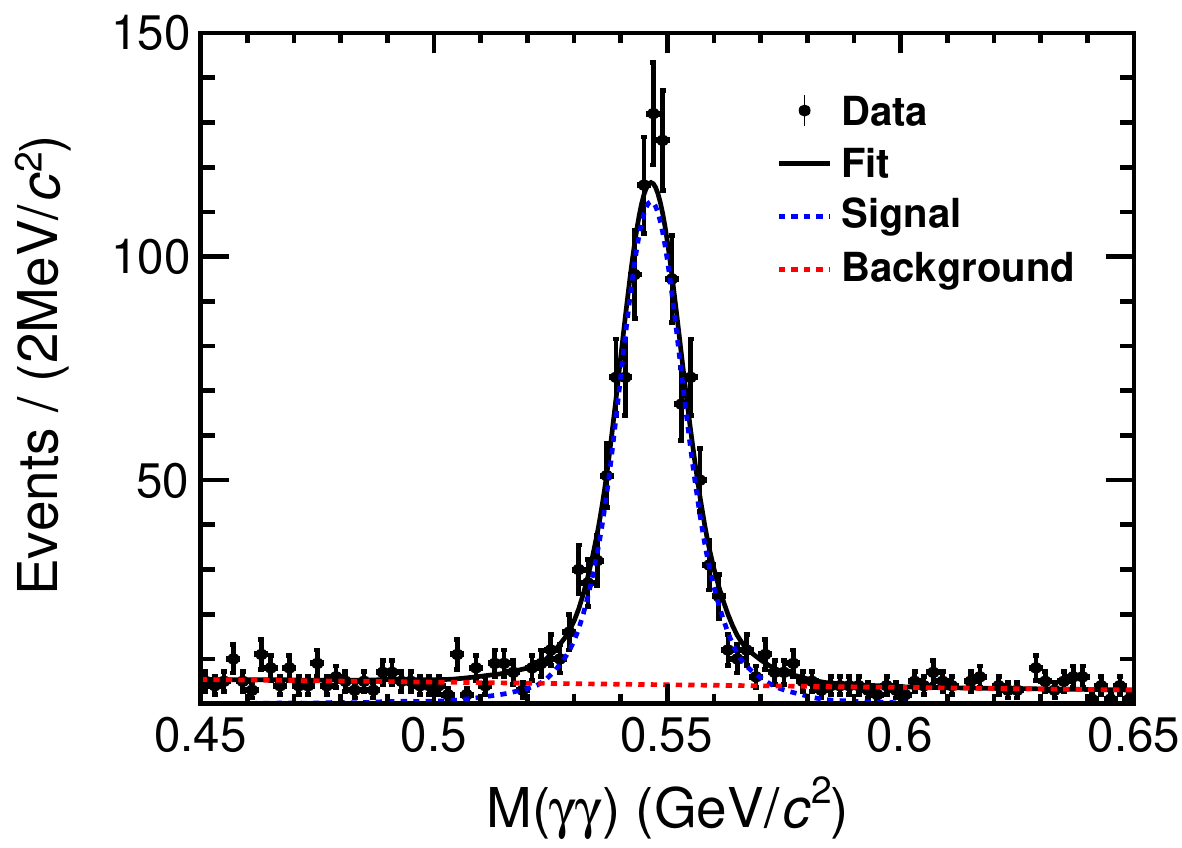}
    \vspace*{-0.3cm}
	\caption{\small Fit to the $M(\gamma\gamma)$ distribution at (left) $\sqrt{s}=2.396$~GeV and (right) $\sqrt{s}=2.900$~GeV. Dots with error bars are data, blue dashed line is the signal shape, red dashed line is the continuum background, the black line gives the total fit result. }
	\label{M2g2396}
\end{figure}
\FloatBarrier

\begin{figure}[!htb]
    \centering
    \includegraphics[width=5.cm]{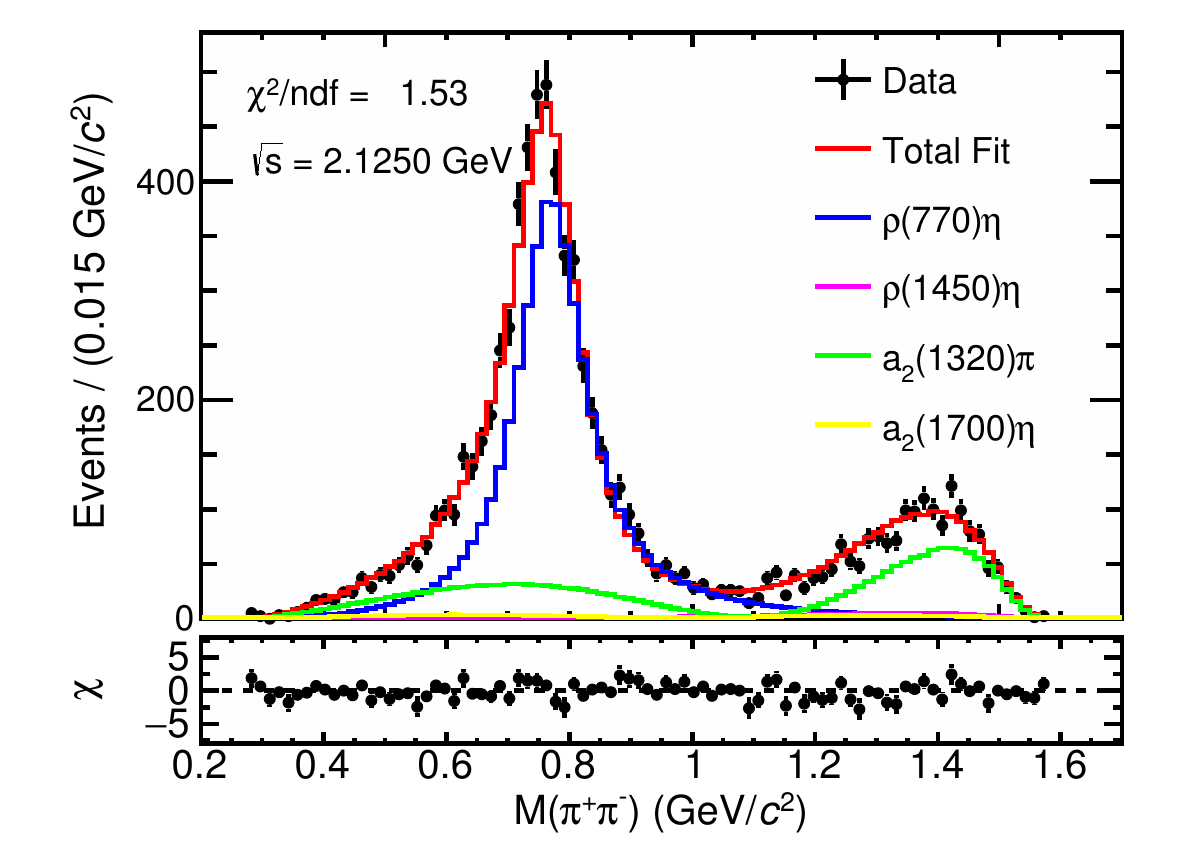}
    \includegraphics[width=5.cm]{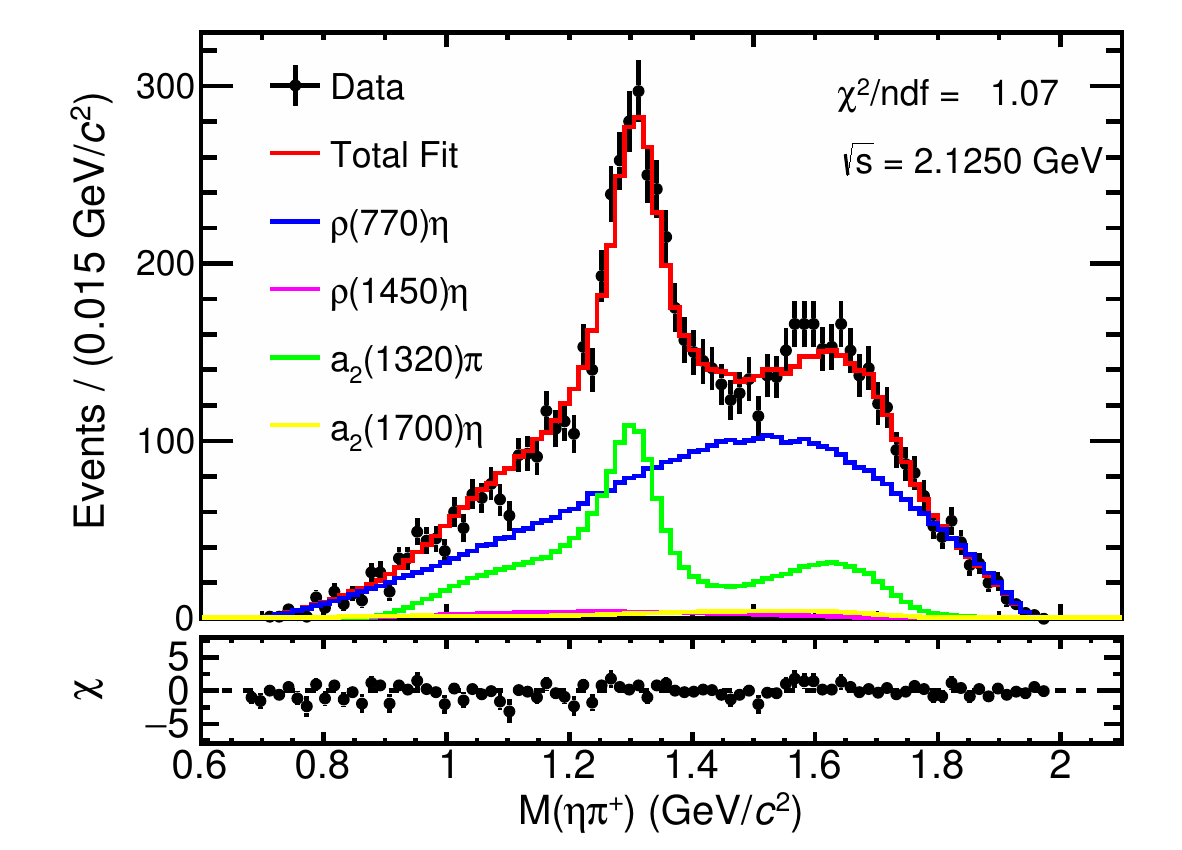}
    \includegraphics[width=5.cm]{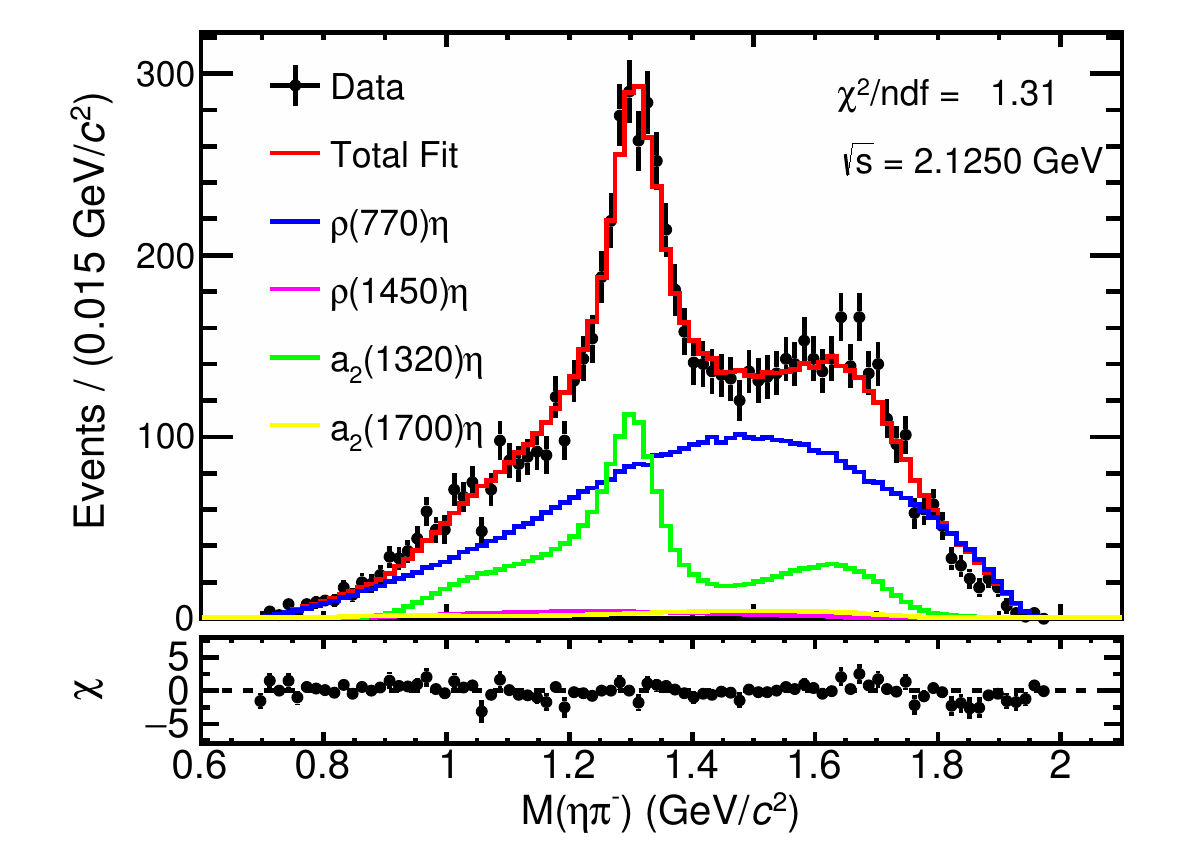}
    \includegraphics[width=7.0cm]{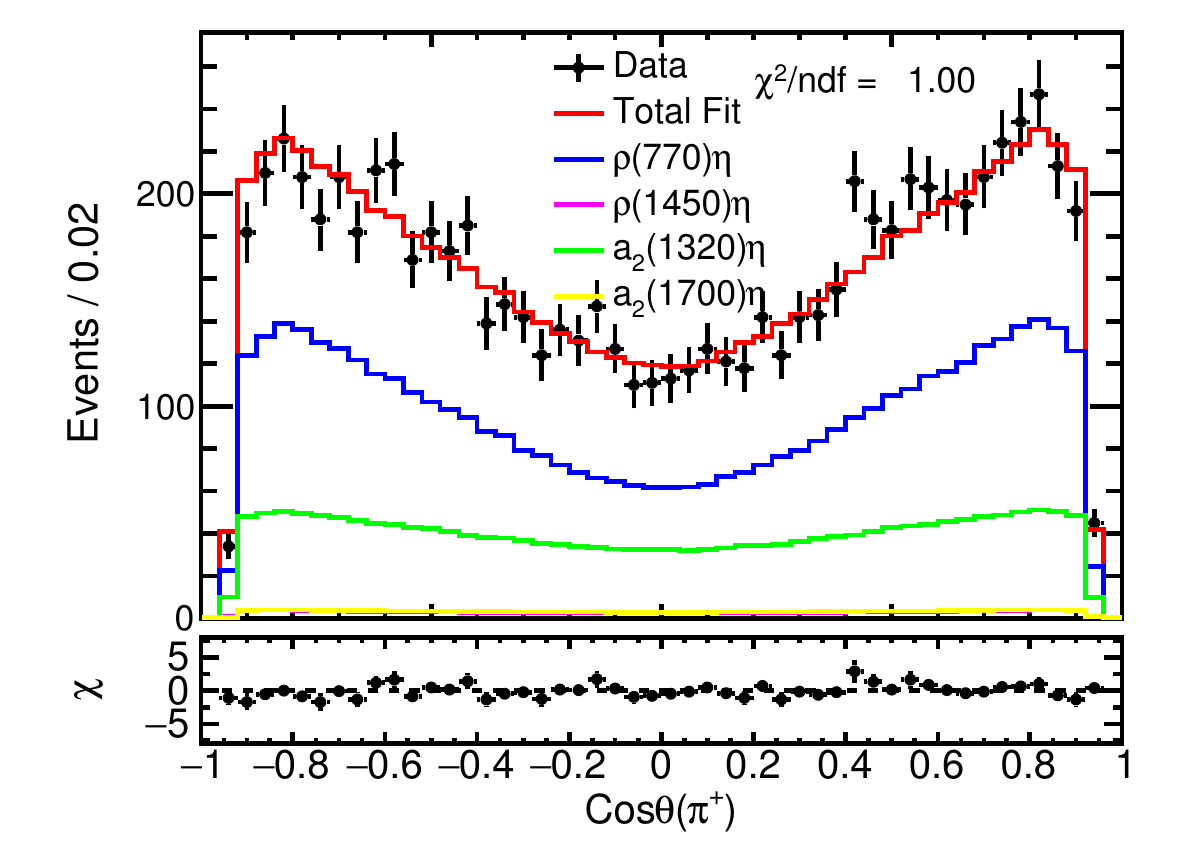}
    \includegraphics[width=7.0cm]{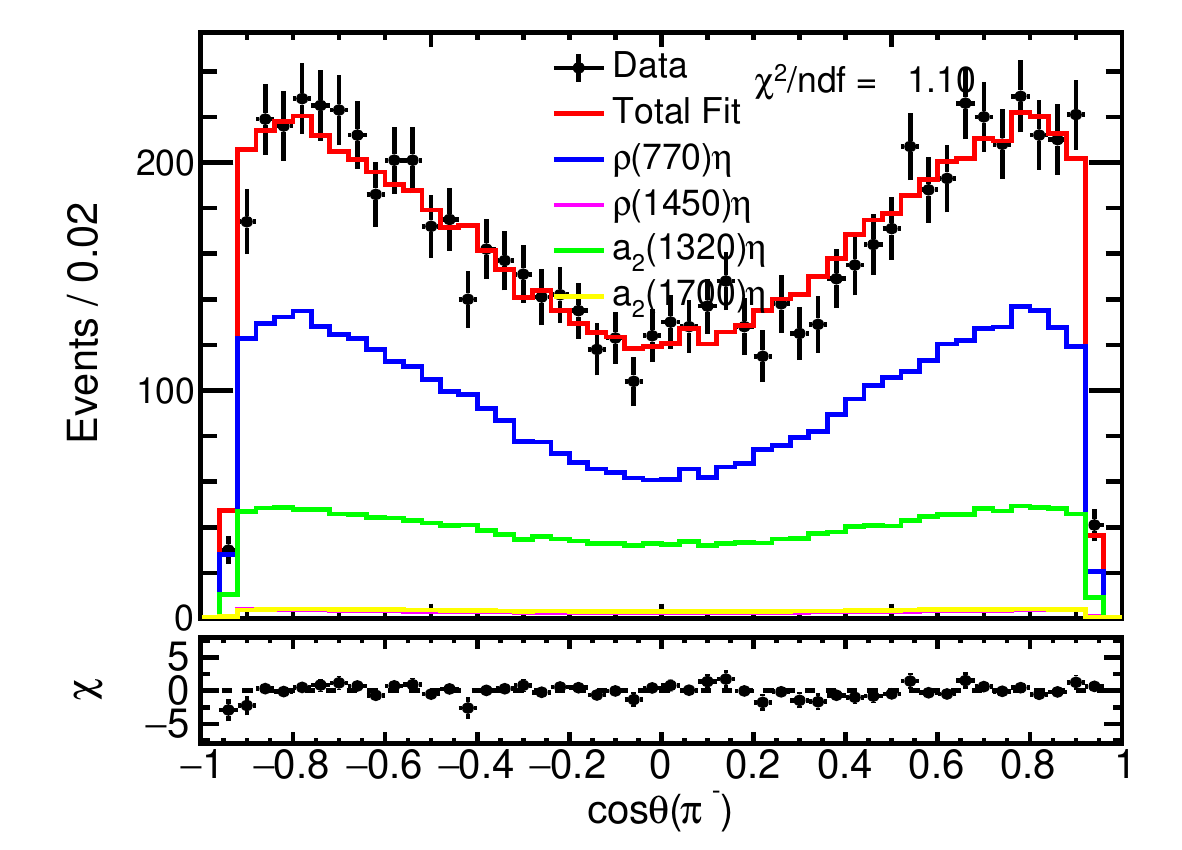}
    \vspace*{-0.3cm}
	\caption{ \small Comparisons of data and PWA fit projections at $\sqrt{s}=2.125$~GeV. Top row: two-body invariant mass distributions. Bottom row: angular distributions.}
	\label{figpwa2125}
\end{figure}
\FloatBarrier

\begin{figure}[!htb]
    \centering
    \includegraphics[width=5.cm]{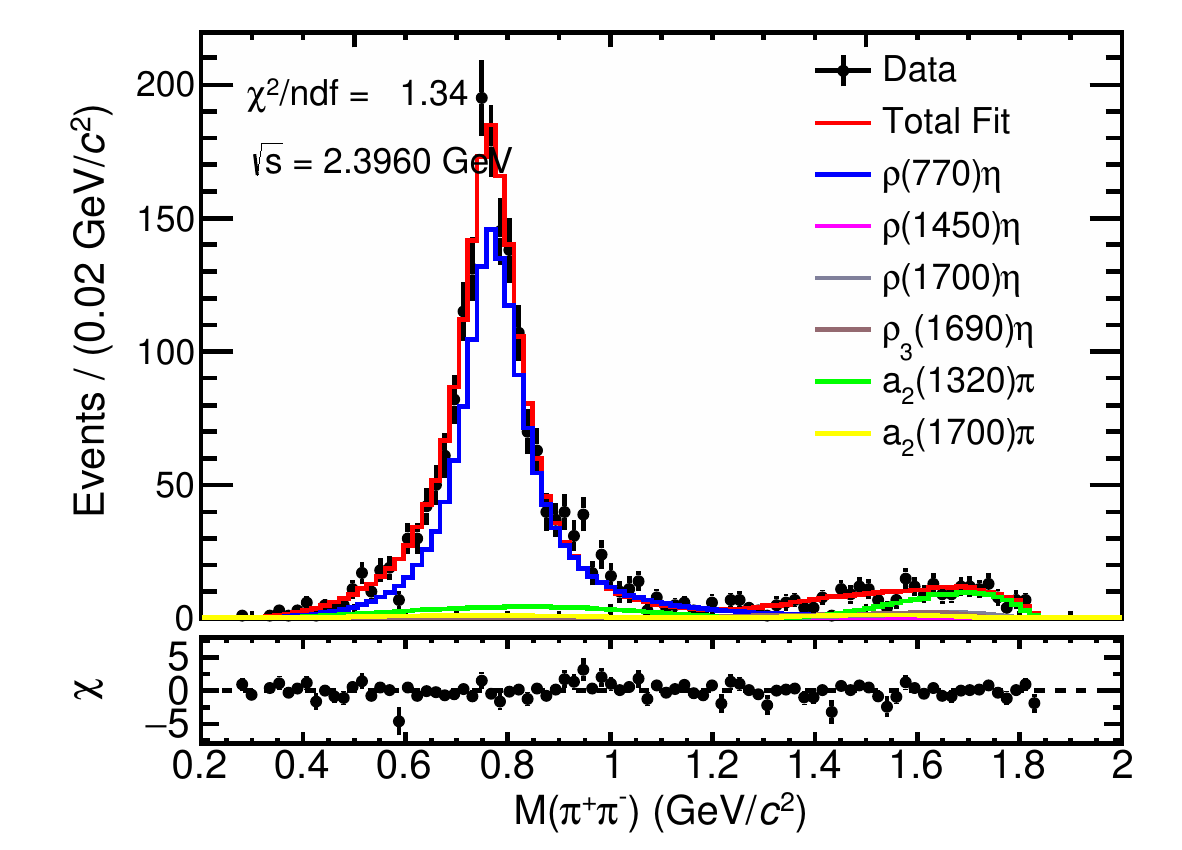}
    \includegraphics[width=5.cm]{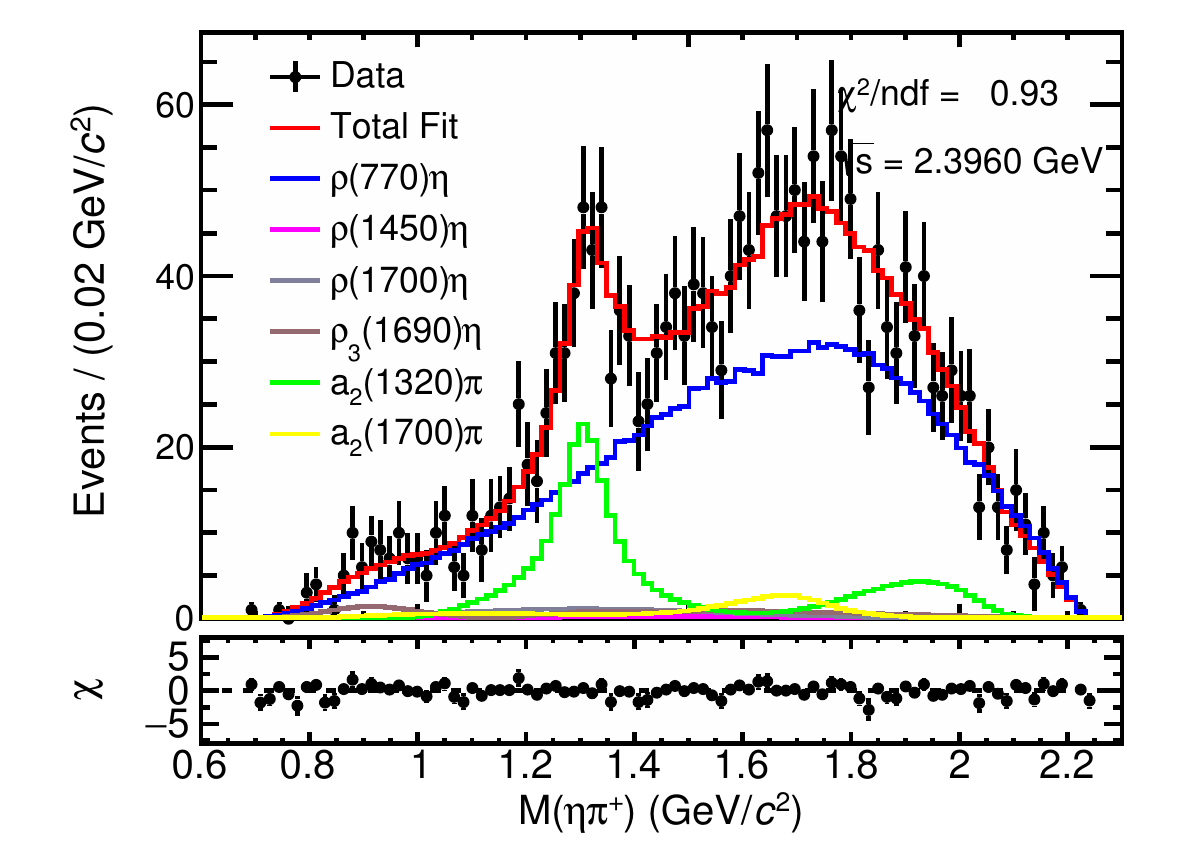}
    \includegraphics[width=5.cm]{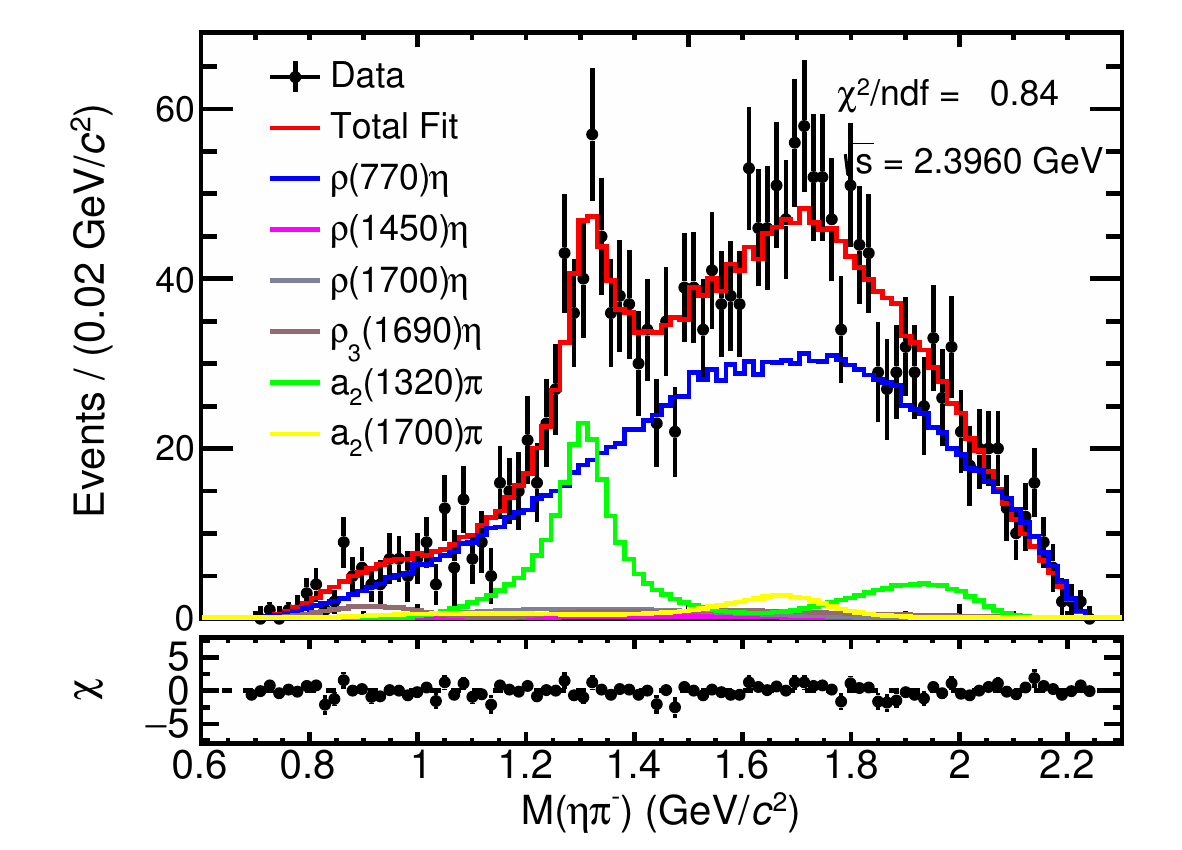}
    \includegraphics[width=7.0cm]{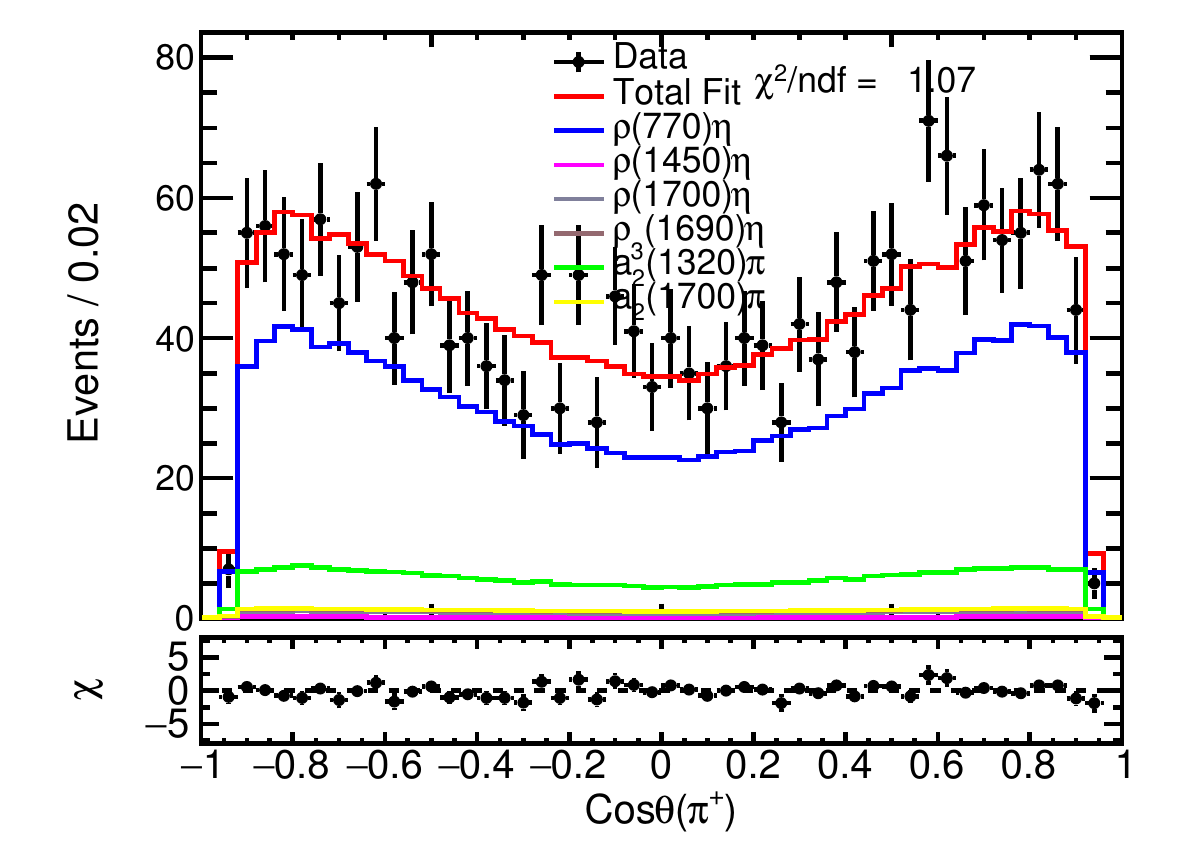}
    \includegraphics[width=7.0cm]{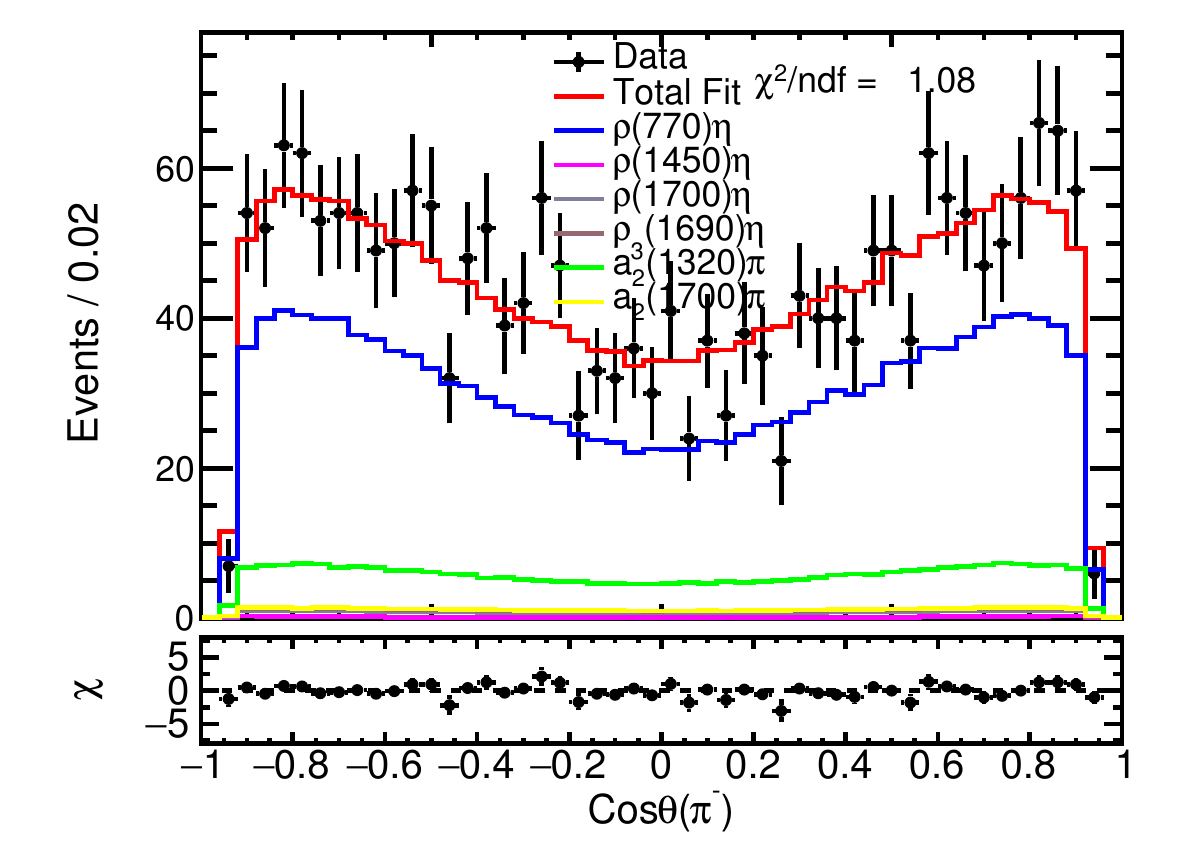}
    \vspace*{-0.3cm}
	\caption{ \small Comparisons of data and PWA fit projections at $\sqrt{s}=2.396$~GeV. Top row: two-body invariant mass distributions. Bottom row: angular distributions.}
	\label{figpwa2396}
\end{figure}
\FloatBarrier

\begin{figure}[!htb]
    \centering
    \includegraphics[width=5.cm]{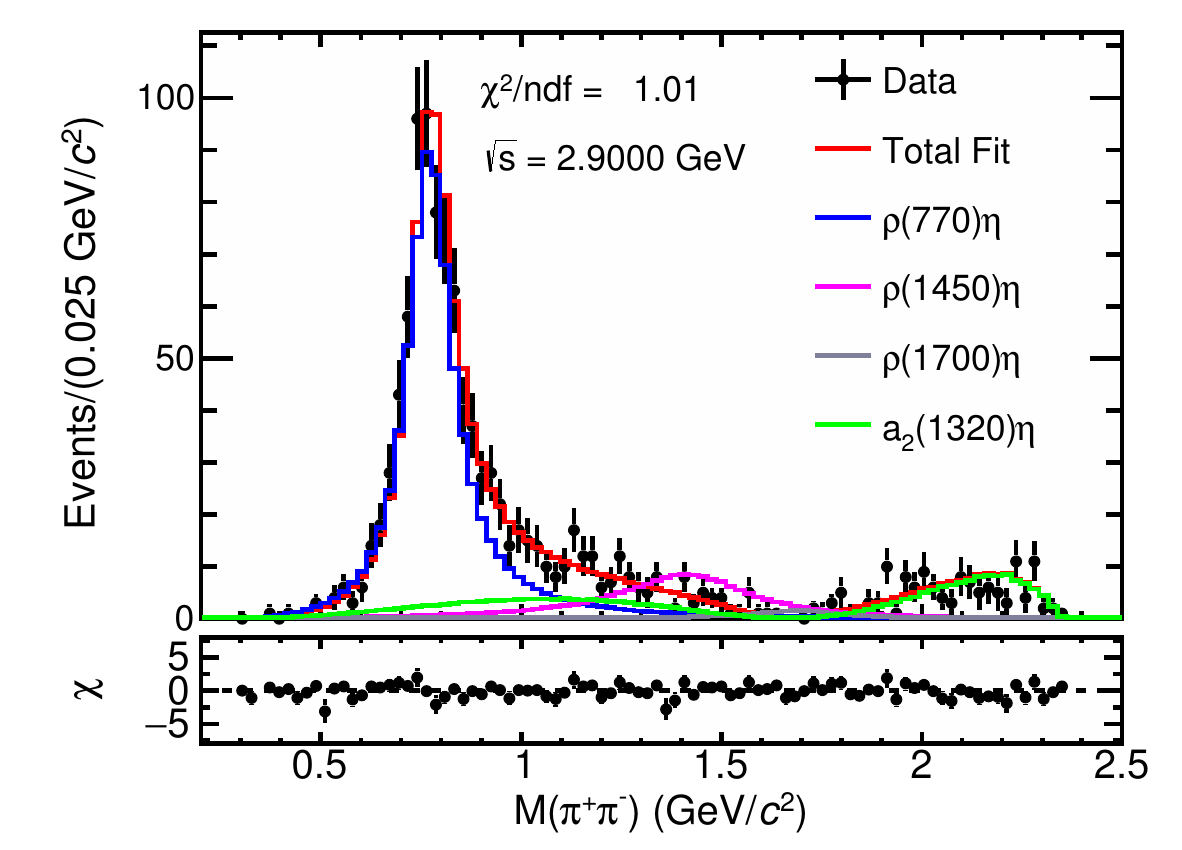}
    \includegraphics[width=5.cm]{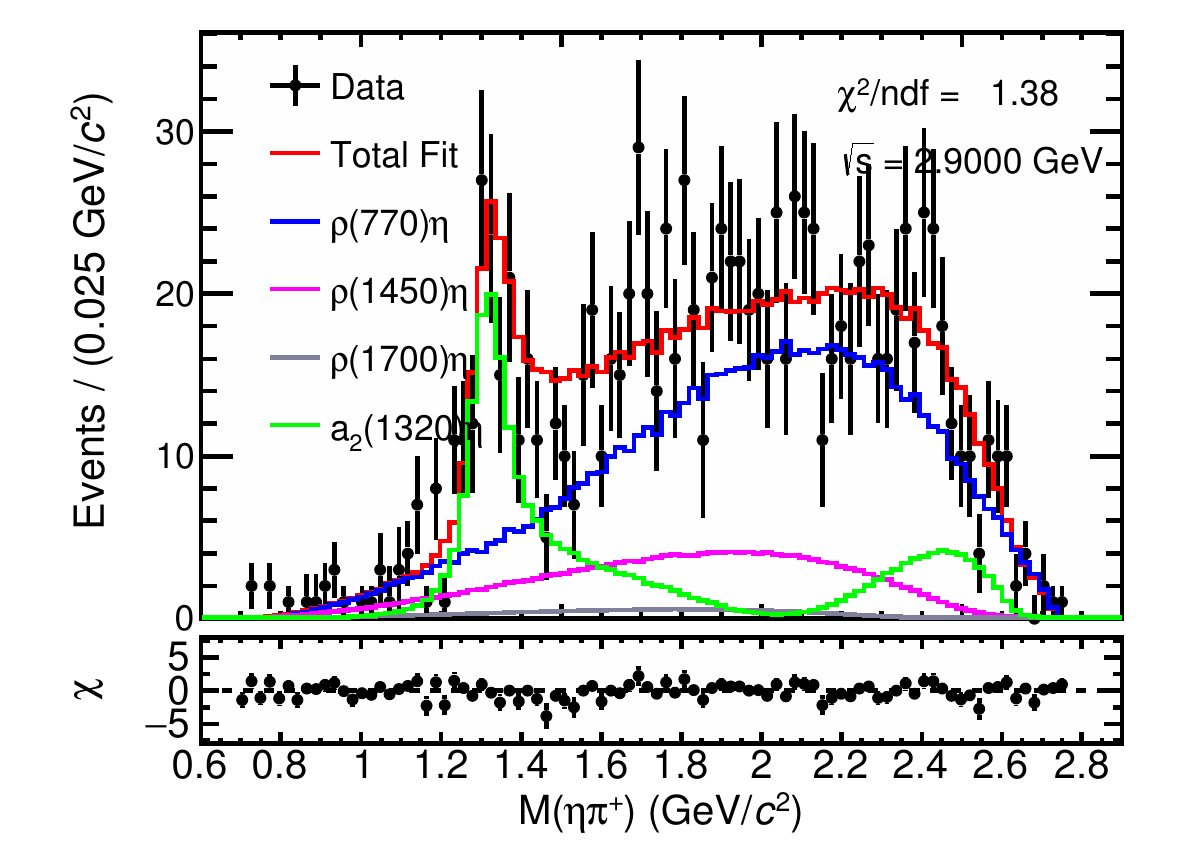}
    \includegraphics[width=5.cm]{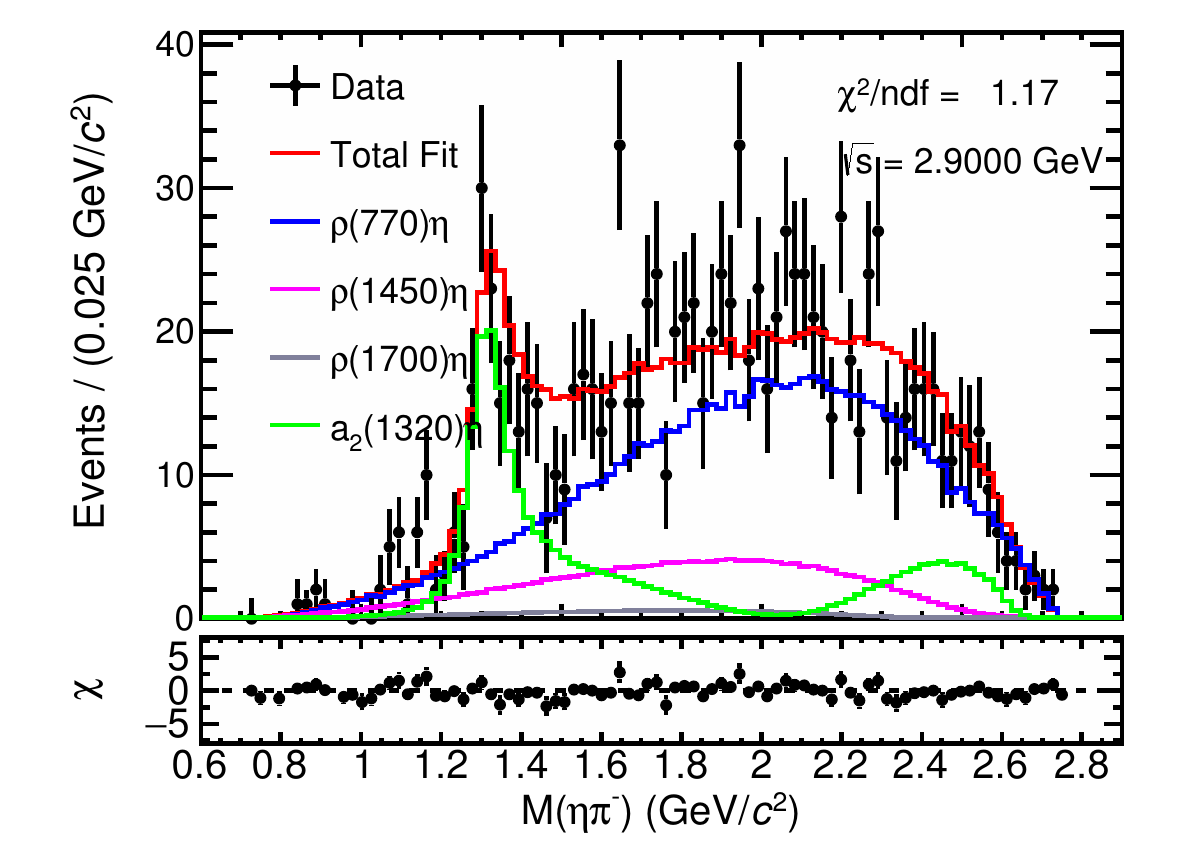}
    \includegraphics[width=7.0cm]{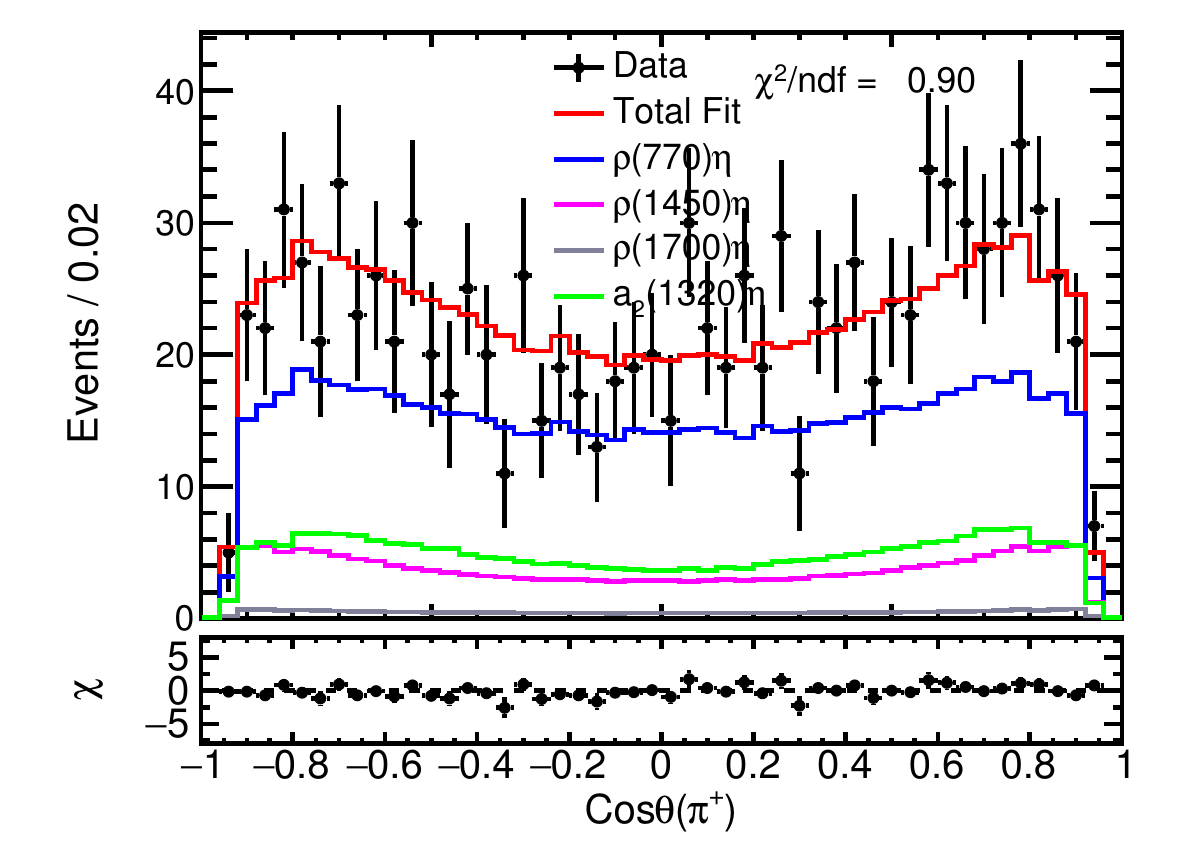}
    \includegraphics[width=7.0cm]{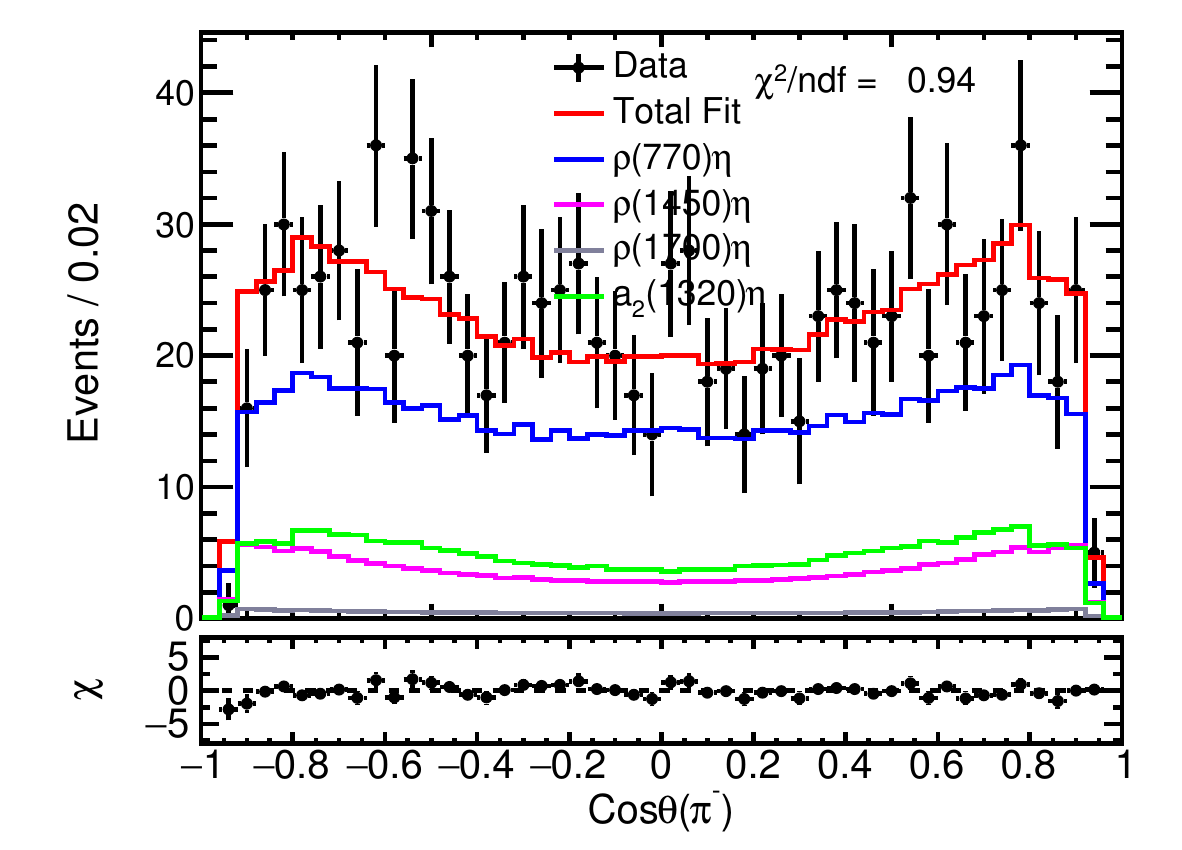}
    \vspace*{-0.3cm}
	\caption{ \small Comparisons of data and PWA fit projections at $\sqrt{s}=2.900$~GeV. Top row: two-body invariant mass distributions. Bottom row: angular distributions.}
	\label{figpwa2900}
\end{figure}
\FloatBarrier

\begin{table*}[!htb]
\begin{center}
\caption{\small The c.m. energy ($\sqrt{s}$), integrated luminosity $\mathcal{L}$,
observed event yields ($N^{\rm obs}$), signal purity (Purity), detection efficiency ($\epsilon$), radiative correction factor $(1+\delta^{\gamma})$, and cross section ($\sigma$) for the process of $\EPP$.
$\frac{1}{|1-\Pi|^2}$ is the vacuum polarization factor taken from QED calculations~\cite{WorkingGrouponRadiativeCorrections:2010bjp}.}
\label{Csetapipi}
\begin{small}
\begin{tabular}{cccccccc}\hline\hline
$\sqrt{s}$~(GeV)   & $\mathcal{L}$~(pb$^{-1}$) &$N^{\rm obs}$ &Purity  & $\epsilon$   & $(1+\delta^{\gamma})$   & $\frac{1}{|1-\Pi|^2}$   & $\sigma$ (pb)   \\
\hline
2.0000   & 10.1   &$1061  \pm33.7$   & 96.4 & 0.264 & 1.14  & 1.04   & $885.7  \pm  28.1 \pm 35.4 $\\
2.0500   & 3.34   &$337.1 \pm19.1$   & 97.1 & 0.264 & 1.15  & 1.04   & $843.5  \pm  47.8 \pm 39.6 $\\
2.1000   & 12.2   &$1040  \pm33.5$   & 94.8 & 0.261 & 1.17  & 1.04   & $708.3  \pm  22.8 \pm 30.5 $\\
2.1250   & 108.   &$8270  \pm94.7$   & 95.6 & 0.257 & 1.19  & 1.04   & $635.3  \pm  7.3  \pm 24.8 $\\
2.1500   & 2.84   &$202.9 \pm14.9$   & 93.6 & 0.248 & 1.25  & 1.04   & $584.8  \pm  42.9 \pm 29.8 $\\
2.1750   & 10.6   &$587.6 \pm25.3$   & 95.4 & 0.242 & 1.30  & 1.04   & $447.1  \pm  19.3 \pm 19.2 $\\
2.2000   & 13.7   &$679.6 \pm27.6$   & 93.5 & 0.234 & 1.32  & 1.04   & $407.5  \pm  16.5 \pm 17.1 $\\
2.2324   & 11.9   &$571.3 \pm25.1$   & 95.9 & 0.235 & 1.32  & 1.04   & $392.7  \pm  17.3 \pm 16.5 $\\
2.3094   & 21.1   &$860.9 \pm30.9$   & 94.3 & 0.234 & 1.32  & 1.04   & $335.2  \pm  12.0 \pm 13.4 $\\
2.3864   & 22.5   &$746.9 \pm28.8$   & 94.1 & 0.232 & 1.34  & 1.04   & $270.9  \pm  10.4 \pm 10.6 $\\
2.3960   & 66.9   &$2182  \pm49.5$   & 93.7 & 0.230 & 1.35  & 1.04   & $266.5  \pm  6.0  \pm 10.4 $\\
2.6444   & 33.7   &$635.1 \pm26.9$   & 92.6 & 0.207 & 1.47  & 1.04   & $157.2  \pm  6.7  \pm 6.4 $\\
2.6464   & 34.0   &$632.2 \pm26.8$   & 93.9 & 0.208 & 1.47  & 1.04   & $154.3  \pm  6.5  \pm 6.2 $\\
2.9000   & 105.   &$1133  \pm36.2$   & 91.0 & 0.183 & 1.65  & 1.03   & $90.7   \pm  2.9  \pm 4.2 $\\
2.9500   & 15.9   &$148.7 \pm13.4$   & 87.3 & 0.177 & 1.69  & 1.03   & $79.3   \pm  7.1  \pm 3.8 $\\
2.9810   & 16.1   &$132.5 \pm12.5$   & 89.3 & 0.174 & 1.72  & 1.03   & $69.8   \pm  6.6  \pm 2.7 $\\
3.0000   & 15.9   &$124.3 \pm12.2$   & 87.6 & 0.173 & 1.74  & 1.02   & $65.9   \pm  6.5  \pm 3.3 $\\
3.0200   & 17.3   &$150.3 \pm13.3$   & 90.1 & 0.168 & 1.76  & 1.02   & $74.6   \pm  6.6  \pm 4.0 $\\
3.0800   & 126.   &$784.7 \pm30.9$   & 86.0 & 0.154 & 1.93  & 0.92   & $53.2   \pm  2.1  \pm 2.2 $\\
\hline \hline
 \end{tabular}
 \end{small}
\end{center}
\end{table*}
\FloatBarrier

\begin{table*}[!htb]
\begin{center}
\caption{\small The c.m. energy ($\sqrt{s}$), integrated luminosity $\mathcal{L}$, observed event yields ($N^{\rm obs}$), detection efficiency ($\epsilon$), radiative correction factor $(1+\delta^{\gamma})$, and cross section ($\sigma$) for the process of $\RHOETA$.
$\frac{1}{|1-\Pi|^2}$ is the vacuum polarization factor taken from QED calculations~\cite{WorkingGrouponRadiativeCorrections:2010bjp}.}
\label{Csrhoeta}
\begin{small}
\begin{tabular}{ccccccc}\hline\hline
$\sqrt{s}$~(GeV)   & $\mathcal{L}$~(pb$^{-1}$) &$N^{\rm obs}$  & $\epsilon$   & $(1+\delta^{\gamma})$   & $\frac{1}{|1-\Pi|^2}$   & $\sigma$ (pb)   \\
\hline
2.0000   & 10.1   &$663.2  \pm 32.9$   & 0.244  & 1.17   & 1.04   & $583.6  \pm  29.0 \pm 35.7 $\\
2.0500   & 3.34   &$173.8  \pm 18.3$   & 0.239  & 1.21   & 1.04   & $456.6  \pm  48.1 \pm 29.2 $\\
2.1000   & 12.2   &$578.2  \pm 33.7$   & 0.237  & 1.23   & 1.04   & $412.5  \pm  24.0 \pm 25.8 $\\
2.1250   & 108.   &$4792.5 \pm 96.1$   & 0.232  & 1.26   & 1.04   & $385.2  \pm  7.7  \pm 23.6 $\\
2.1500   & 2.84   &$122.4  \pm 13.8$   & 0.226  & 1.29   & 1.04   & $375.1  \pm  42.3 \pm 25.7 $\\
2.1750   & 10.6   &$332.5  \pm 26.0$   & 0.216  & 1.40   & 1.04   & $263.2  \pm  20.6 \pm 16.2 $\\
2.2000   & 13.7   &$382.6  \pm 26.0$   & 0.207  & 1.40   & 1.04   & $244.5  \pm  16.6 \pm 15.4 $\\
2.2324   & 11.9   &$376.9  \pm 27.3$   & 0.230  & 1.28   & 1.04   & $273.0  \pm  19.8 \pm 17.2 $\\
2.3094   & 21.1   &$620.8  \pm 43.8$   & 0.224  & 1.32   & 1.04   & $252.5  \pm  17.8 \pm 15.3 $\\
2.3864   & 22.5   &$441.7  \pm 36.4$   & 0.217  & 1.37   & 1.04   & $167.6  \pm  13.8 \pm 10.1 $\\
2.3960   & 66.9   &$1516.3 \pm 64.1$   & 0.217  & 1.37   & 1.04   & $193.5  \pm  8.2  \pm 11.8 $\\
2.6444   & 33.7   &$426.0  \pm 29.1$   & 0.197  & 1.52   & 1.04   & $107.1  \pm  7.3  \pm 6.6  $\\
2.6464   & 34.0   &$404.3  \pm 25.8$   & 0.196  & 1.52   & 1.04   & $101.3  \pm  6.5  \pm 6.0  $\\
2.9000   & 105.   &$757.1  \pm 34.7$   & 0.177  & 1.71   & 1.03   & $60.4   \pm  2.8  \pm 8.2  $\\
2.9500   & 15.9   &$98.7   \pm 12.2$   & 0.172  & 1.75   & 1.03   & $52.3   \pm  6.5  \pm 7.1  $\\
2.9810   & 16.1   &$83.7   \pm 10.9$   & 0.173  & 1.76   & 1.03   & $43.3   \pm  5.6  \pm 5.8  $\\
3.0000   & 15.9   &$84.1   \pm 12.8$   & 0.170  & 1.78   & 1.02   & $44.4   \pm  6.8  \pm 6.1  $\\
3.0200   & 17.3   &$92.3   \pm 11.3$   & 0.168  & 1.79   & 1.02   & $45.0   \pm  5.5  \pm 6.2  $\\
3.0800   & 126.   &$517.4  \pm 31.1$   & 0.153  & 1.98   & 0.92   & $34.4   \pm  2.1  \pm 4.6  $\\
\hline \hline
 \end{tabular}
 \end{small}
\end{center}
\end{table*}
\FloatBarrier

\begin{table*}[!htb]
\begin{center}
\caption{\small The c.m. energy ($\sqrt{s}$), integrated luminosity $\mathcal{L}$, observed event yields ($N^{\rm obs}$), detection efficiency ($\epsilon$), radiative correction factor $(1+\delta^{\gamma})$, and cross section ($\sigma$) for the process of $\API$.
$\frac{1}{|1-\Pi|^2}$ is the vacuum polarization factor taken from QED calculations~\cite{WorkingGrouponRadiativeCorrections:2010bjp}.}
\label{Csapi}
\begin{small}
\begin{tabular}{ccccccc}\hline\hline
$\sqrt{s}$~(GeV)   & $\mathcal{L}$~(pb$^{-1}$) &$N^{\rm obs}$  & $\epsilon$   & $(1+\delta^{\gamma})$   & $\frac{1}{|1-\Pi|^2}$   & $\sigma$ (pb)   \\
\hline
2.0000   & 10.1   &$257.8  \pm 19.8$   & 0.335  & 0.90   & 1.04   & $1482.6  \pm  113.9 \pm 188.2$ \\
2.0500   & 3.34   &$115.0  \pm 19.3$   & 0.329  & 0.92   & 1.04   & $1992.2  \pm  334.3 \pm 255.6$ \\
2.1000   & 12.2   &$301.7  \pm 28.7$   & 0.313  & 0.98   & 1.04   & $1411.9  \pm  134.3 \pm 179.8$ \\
2.1250   & 108.   &$1993.9 \pm 72.9$   & 0.303  & 1.02   & 1.04   & $1046.2  \pm  38.2  \pm 132.7$ \\
2.1500   & 2.84   &$64.5   \pm 10.4$   & 0.289  & 1.07   & 1.04   & $1286.2  \pm  207.4 \pm 168.4$ \\
2.1750   & 10.6   &$123.6  \pm 17.1$   & 0.279  & 1.11   & 1.04   & $659.4   \pm  91.2  \pm 83.7 $ \\
2.2000   & 13.7   &$167.0  \pm 19.2$   & 0.268  & 1.16   & 1.04   & $686.7   \pm  78.9  \pm 87.7 $ \\
2.2324   & 11.9   &$82.0   \pm 18.0$   & 0.253  & 1.21   & 1.04   & $394.2   \pm  86.5  \pm 50.4 $ \\
2.3094   & 21.1   &$176.6  \pm 23.2$   & 0.239  & 1.26   & 1.04   & $486.7   \pm  63.9  \pm 71.5 $ \\
2.3864   & 22.5   &$98.1   \pm 17.2$   & 0.237  & 1.24   & 1.04   & $259.8   \pm  45.6  \pm 38.1 $ \\
2.3960   & 66.9   &$283.4  \pm 25.5$   & 0.238  & 1.24   & 1.04   & $251.4   \pm  22.6  \pm 36.8 $ \\
2.6444   & 33.7   &$107.7  \pm 11.4$   & 0.234  & 1.17   & 1.04   & $204.4   \pm  21.6  \pm 30.1 $ \\
2.6464   & 34.0   &$110.1  \pm 13.7$   & 0.236  & 1.17   & 1.04   & $205.4   \pm  25.6  \pm 30.0 $ \\
2.9000   & 105.   &$245.4  \pm 24.6$   & 0.223  & 1.17   & 1.03   & $156.9   \pm  15.7  \pm 23.3 $ \\
2.9500   & 15.9   &$26.1   \pm 8.6 $   & 0.216  & 1.16   & 1.03   & $114.7   \pm  37.8  \pm 17.2 $ \\
2.9810   & 16.1   &$35.0   \pm 6.7 $   & 0.218  & 1.15   & 1.03   & $151.9   \pm  29.1  \pm 22.4 $ \\
3.0000   & 15.9   &$28.4   \pm 9.7 $   & 0.214  & 1.16   & 1.02   & $126.0   \pm  43.0  \pm 19.0 $ \\
3.0200   & 17.3   &$43.6   \pm 7.7 $   & 0.215  & 1.16   & 1.02   & $177.0   \pm  31.3  \pm 26.9 $ \\
3.0800   & 126.   &$196.9  \pm 15.2$   & 0.203  & 1.21   & 0.92   & $111.4   \pm  8.6   \pm 16.4 $ \\
\hline \hline
 \end{tabular}
 \end{small}
\end{center}
\end{table*}
\FloatBarrier

\begin{table*}[!htb]
\caption{Relative systematic uncertainties (in \%) from luminosity (Lum), tracking, photon efficiency, PID, MC model, kinematic fit (KF), radiative correction (RC) due to ISR, $E/p$ ratio, helicity angle (HA), signal yields (SY), cited branching fraction (Br), for the measured cross sections of $\EPP$. The sources with an superscript ``$*$'' are common
relative systematic uncertainties for different c.m. energies.
They are considered to be 100\% correlated among c.m. energies. }
\begin{center}
\begin{tabular}{ccccccccccccc}\hline\hline
$\sqrt{s}$~(GeV) &Lum$^*$ &Tracking$^*$  &Photon$^*$ &PID$^*$ &MC &KF &RC &E/p$^*$  &HA$^*$    &SY     &Br$^*$ &Total\\ \hline
2.0000    &1.0  &2.0    &2.0    &2.0    &0.9   & 0.4   &1.0    &0.7    &0.2    &0.4    &0.5    &4.0   \\
2.0500    &1.0  &2.0    &2.0    &2.0    &1.7   & 0.4   &1.3    &0.7    &0.2    &1.9    &0.5    &4.7   \\
2.1000    &1.0  &2.0    &2.0    &2.0    &1.1   & 0.5   &1.5    &0.7    &0.2    &1.1    &0.5    &4.3   \\
2.1250    &1.0  &2.0    &2.0    &2.0    &0.2   & 0.5   &1.1    &0.7    &0.2    &0.1    &0.5    &3.9   \\
2.1500    &1.0  &2.0    &2.0    &2.0    &0.5   & 0.5   &1.2    &0.7    &0.2    &3.1    &0.5    &5.1   \\
2.1750    &1.0  &2.0    &2.0    &2.0    &1.6   & 0.5   &1.3    &0.7    &0.2    &0.4    &0.5    &4.3   \\
2.2000    &1.0  &2.0    &2.0    &2.0    &0.5   & 0.4   &1.2    &0.7    &0.2    &1.5    &0.5    &4.2   \\
2.2324    &1.0  &2.0    &2.0    &2.0    &0.6   & 0.4   &1.1    &0.7    &0.2    &1.5    &0.5    &4.2   \\
2.3094    &1.0  &2.0    &2.0    &2.0    &0.5   & 0.4   &0.8    &0.7    &0.2    &1.0    &0.5    &4.0   \\
2.3864    &1.0  &2.0    &2.0    &2.0    &0.8   & 0.4   &0.7    &0.7    &0.2    &0.6    &0.5    &3.9   \\
2.3960    &1.0  &2.0    &2.0    &2.0    &0.2   & 0.5   &0.8    &0.7    &0.2    &0.8    &0.5    &3.9   \\
2.6444    &1.0  &2.0    &2.0    &2.0    &0.1   & 0.4   &0.6    &0.7    &0.2    &1.4    &0.5    &4.1   \\
2.6464    &1.0  &2.0    &2.0    &2.0    &1.2   & 0.5   &0.6    &0.7    &0.2    &0.2    &0.5    &4.0   \\
2.9000    &1.0  &2.0    &2.0    &2.0    &0.8   & 0.5   &0.7    &0.7    &0.2    &2.4    &0.5    &4.6   \\
2.9500    &1.0  &2.0    &2.0    &2.0    &0.3   & 0.5   &0.7    &0.7    &0.2    &2.9    &0.5    &4.8   \\
2.9810    &1.0  &2.0    &2.0    &2.0    &0.4   & 0.4   &0.7    &0.7    &0.2    &1.0    &0.5    &3.9   \\
3.0000    &1.0  &2.0    &2.0    &2.0    &0.4   & 0.5   &0.7    &0.7    &0.2    &3.2    &0.5    &5.0   \\
3.0200    &1.0  &2.0    &2.0    &2.0    &0.1   & 0.5   &0.7    &0.7    &0.2    &3.8    &0.5    &5.4   \\
3.0800    &1.0  &2.0    &2.0    &2.0    &1.5   & 0.5   &0.7    &0.7    &0.2    &0.9    &0.5    &4.2   \\
\hline\hline
\end{tabular}
\label{SYSetapipi}
\end{center}
\end{table*}
\FloatBarrier

\begin{table}[!htb]
\caption{Relative systematic uncertainties (in \%) from luminosity (Lum), tracking, photon efficiency, PID, kinematic fit (KF), radiative correction (RC) due to ISR, $E/p$ ratio, helicity angle (HA), signal yields (SY), cited branching fraction (Br), Barrier factor (BF), resonance parameters (Res. para.), BW parametrization (BW para.), background estimation (Bkg) and extra resonances (Ext), for the measured cross sections of $\RHOETA$. The sources with an upscript $*$ are common for different c.m. energies.}

\begin{center}
\begin{tabular}{ccccccccccccccccc}\hline\hline
$\sqrt{s}$~(GeV) &Lum$^*$ &Tracking$^*$ &Photon$^*$ &PID$^*$ &KF &RC &E/p$^*$  &HA$^*$  &SY  &Br$^*$ &BF &Res.para. &BW para. &Bkg &Ext.res &Total\\ \hline
2.0000   &1.0   &2.0    &2.0    &2.0    &0.5   &0.6   &0.7   &0.2   &0.4   &0.5   &0.8   &1.6   &0.6   &0.3   &4.4    &6.1   \\
2.0500   &1.0   &2.0    &2.0    &2.0    &0.5   &0.5   &0.7   &0.2   &1.9   &0.5   &0.8   &1.6   &0.6   &0.3   &4.4    &6.4   \\
2.1000   &1.0   &2.0    &2.0    &2.0    &0.5   &1.1   &0.7   &0.2   &1.1   &0.5   &0.8   &1.6   &0.6   &0.3   &4.4    &6.3  \\
2.1250   &1.0   &2.0    &2.0    &2.0    &0.5   &0.8   &0.7   &0.2   &0.1   &0.5   &0.8   &1.6   &0.6   &0.3   &4.4    &6.1   \\
2.1500   &1.0   &2.0    &2.0    &2.0    &0.6   &0.5   &0.7   &0.2   &3.1   &0.5   &0.8   &1.6   &0.6   &0.3   &4.4    &6.9   \\
2.1750   &1.0   &2.0    &2.0    &2.0    &0.7   &0.8   &0.7   &0.2   &0.4   &0.5   &0.8   &1.6   &0.6   &0.3   &4.4    &6.1   \\
2.2000   &1.0   &2.0    &2.0    &2.0    &0.6   &0.9   &0.7   &0.2   &1.5   &0.5   &0.8   &1.6   &0.6   &0.3   &4.4    &6.3   \\
2.2324   &1.0   &2.0    &2.0    &2.0    &0.4   &0.9   &0.7   &0.2   &1.5   &0.5   &0.8   &1.6   &0.6   &0.3   &4.4    &6.3   \\
2.3094   &1.0   &2.0    &2.0    &2.0    &0.5   &0.5   &0.7   &0.2   &1.0   &0.5   &3.2   &1.7   &1.2   &1.3   &2.3    &6.1   \\
2.3864   &1.0   &2.0    &2.0    &2.0    &0.5   &0.9   &0.7   &0.2   &0.6   &0.5   &3.2   &1.7   &1.2   &1.3   &2.3    &6.0   \\
2.3960   &1.0   &2.0    &2.0    &2.0    &0.4   &1.1   &0.7   &0.2   &0.8   &0.5   &3.2   &1.7   &1.2   &1.3   &2.3    &6.1   \\
2.6444   &1.0   &2.0    &2.0    &2.0    &0.6   &0.5   &0.7   &0.2   &1.4   &0.5   &3.2   &1.7   &1.2   &1.3   &2.3    &6.1   \\
2.6464   &1.0   &2.0    &2.0    &2.0    &0.5   &0.5   &0.7   &0.2   &0.2   &0.5   &3.2   &1.7   &1.2   &1.3   &2.3    &6.0   \\
2.9000   &1.0   &2.0    &2.0    &2.0    &0.5   &0.9   &0.7   &0.2   &2.4   &0.5   &2.1   &2.8   &1.3   &4.6   &11.3   &13.6  \\
2.9500   &1.0   &2.0    &2.0    &2.0    &0.4   &1.0   &0.7   &0.2   &2.9   &0.5   &2.1   &2.8   &1.3   &4.6   &11.3   &13.7  \\
2.9810   &1.0   &2.0    &2.0    &2.0    &0.3   &0.7   &0.7   &0.2   &1.0   &0.5   &2.1   &2.8   &1.3   &4.6   &11.3   &13.4  \\
3.0000   &1.0   &2.0    &2.0    &2.0    &0.5   &1.0   &0.7   &0.2   &3.2   &0.5   &2.1   &2.8   &1.3   &4.6   &11.3   &13.7  \\
3.0200   &1.0   &2.0    &2.0    &2.0    &0.3   &0.9   &0.7   &0.2   &3.8   &0.5   &2.1   &2.8   &1.3   &4.6   &11.3   &13.9  \\
3.0800   &1.0   &2.0    &2.0    &2.0    &0.4   &0.5   &0.7   &0.2   &0.9   &0.5   &2.1   &2.8   &1.3   &4.6   &11.3   &13.3  \\
\hline\hline
\end{tabular}
\label{tab:sysy_rhoeta}
\end{center}
\end{table}
\FloatBarrier

\begin{table}[!htb]
\caption{Relative systematic uncertainties (in \%) from luminosity (Lum), tracking, photon efficiency, PID, kinematic fit (KF), radiative correction (RC) due to ISR, $E/p$ ratio, helicity angle (HA), signal yields (SY), cited branching fraction (Br), Barrier factor (BF), resonance parameters (Res. para.), BW parametrization (BW para.), background estimation (Bkg) and extra resonances (Ext), for the measured cross sections of $\API$. The sources with an upscript $*$ are common for different c.m. energies.}
\begin{center}
\begin{tabular}{ccccccccccccccccc}\hline\hline
$\sqrt{s}$~(GeV) &Lum$^*$ &Tracking$^*$ &Photon$^*$ &PID$^*$ &KF &RC &E/p$^*$ &HA$^*$ &SY &Br$^*$ &BF &Res.para &BW para. &Bkg &Ext.res &Total\\\hline
2.0000   &1.0   &2.0    &2.0    &2.0    &0.4     &0.7     &0.7     &0.2     &0.4     &8.3   &3.6   &1.6   &5.5   &3.0   &4.9    &12.7  \\
2.0500   &1.0   &2.0    &2.0    &2.0    &0.5     &0.5     &0.7     &0.2     &1.9     &8.3   &3.6   &1.6   &5.5   &3.0   &4.9    &12.8  \\
2.1000   &1.0   &2.0    &2.0    &2.0    &0.5     &0.7     &0.7     &0.2     &1.1     &8.3   &3.6   &1.6   &5.5   &3.0   &4.9    &12.7  \\
2.1250   &1.0   &2.0    &2.0    &2.0    &0.5     &0.6     &0.7     &0.2     &0.1     &8.3   &3.6   &1.6   &5.5   &3.0   &4.9    &12.7  \\
2.1500   &1.0   &2.0    &2.0    &2.0    &0.5     &0.9     &0.7     &0.2     &3.1     &8.3   &3.6   &1.6   &5.5   &3.0   &4.9    &13.1  \\
2.1750   &1.0   &2.0    &2.0    &2.0    &0.5     &0.5     &0.7     &0.2     &0.4     &8.3   &3.6   &1.6   &5.5   &3.0   &4.9    &12.7  \\
2.2000   &1.0   &2.0    &2.0    &2.0    &0.5     &0.6     &0.7     &0.2     &1.5     &8.3   &3.6   &1.6   &5.5   &3.0   &4.9    &12.8  \\
2.2324   &1.0   &2.0    &2.0    &2.0    &0.5     &0.6     &0.7     &0.2     &1.5     &8.3   &3.6   &1.6   &5.5   &3.0   &4.9    &12.8  \\
2.3094   &1.0   &2.0    &2.0    &2.0    &0.5     &0.9     &0.7     &0.2     &1.0     &8.3   &5.1   &2.2   &3.3   &1.6   &9.3    &14.7  \\
2.3864   &1.0   &2.0    &2.0    &2.0    &0.4     &0.8     &0.7     &0.2     &0.6     &8.3   &5.1   &2.2   &3.3   &1.6   &9.3    &14.7  \\
2.3960   &1.0   &2.0    &2.0    &2.0    &0.5     &0.6     &0.7     &0.2     &0.8     &8.3   &5.1   &2.2   &3.3   &1.6   &9.3    &14.7  \\
2.6444   &1.0   &2.0    &2.0    &2.0    &0.5     &1.1     &0.7     &0.2     &1.4     &8.3   &5.1   &2.2   &3.3   &1.6   &9.3    &14.7  \\
2.6464   &1.0   &2.0    &2.0    &2.0    &0.4     &0.6     &0.7     &0.2     &0.2     &8.3   &5.1   &2.2   &3.3   &1.6   &9.3    &14.6  \\
2.9000   &1.0   &2.0    &2.0    &2.0    &0.3     &0.6     &0.7     &0.2     &2.4     &8.3   &4.4   &2.2   &1.3   &3.9   &9.6    &14.9  \\
2.9500   &1.0   &2.0    &2.0    &2.0    &0.4     &0.5     &0.7     &0.2     &2.9     &8.3   &4.4   &2.2   &1.3   &3.9   &9.6    &15.0  \\
2.9810   &1.0   &2.0    &2.0    &2.0    &0.5     &1.0     &0.7     &0.2     &1.0     &8.3   &4.4   &2.2   &1.3   &3.9   &9.6    &14.7  \\
3.0000   &1.0   &2.0    &2.0    &2.0    &0.5     &1.0     &0.7     &0.2     &3.2     &8.3   &4.4   &2.2   &1.3   &3.9   &9.6    &15.0  \\
3.0200   &1.0   &2.0    &2.0    &2.0    &0.3     &0.9     &0.7     &0.2     &3.8     &8.3   &4.4   &2.2   &1.3   &3.9   &9.6    &15.2  \\
3.0800   &1.0   &2.0    &2.0    &2.0    &0.5     &0.6     &0.7     &0.2     &0.9     &8.3   &4.4   &2.2   &1.3   &3.9   &9.6    &14.7  \\
\hline\hline
\end{tabular}
\label{tab:sysy_a2pi}
\end{center}
\end{table}
\FloatBarrier

\bibliography{reference}